\documentclass[preprints,article,accept,pdftex,moreauthors]{Definitions/mdpi} 

\usepackage{bm}
\usepackage{graphicx,color}
\usepackage{amsmath,amssymb}
\usepackage{feynmf}
\usepackage{graphicx}
\graphicspath{{Fig/}}
\newcommand{\besc}{\begin{screen} \vspace{-0.35cm}}
\newcommand{\eesc}{ \end{screen} \\}
\newcommand{\bft}{\vspace{0.2cm} \begin{footnotesize} }
\newcommand{\eft}{ \end{footnotesize} }
\newcommand{\beq}{\begin{equation}}
\newcommand{\eeq}{\end{equation}}

\newcommand{\tr}{{\rm tr}}
\newcommand{\Tr}{{\rm Tr}}

\newcommand{\lqcd}{\Lambda_{\rm QCD}}
\newcommand{\bp}{\bm{p}}
\newcommand{\bq}{\bm{q}}

\newcommand{\br}{\bm{r}}

\newcommand{\bK}{\bm{K}}

\newcommand{\bR}{\bm{R}}

\newcommand{\brho}{\bm{\rho}}
\newcommand{\blambda}{\bm{\lambda}}
\newcommand{\la}{\langle}
\newcommand{\ra}{\rangle}

\newcommand{\calS}{\mathcal{S}}

\newcommand{\calN}{\mathcal{N}}

\newcommand{\calK}{\mathcal{K}}

\newcommand{\calP}{\mathcal{P}}
\newcommand{\calW}{\mathcal{W}}

\newcommand{\calY}{\mathcal{Y}}
\newcommand{\rmd}{\mathrm{d}}
\newcommand{\rmi}{\mathrm{i}}
\newcommand{\rme}{\mathrm{e}}

\newcommand{\rmB}{\mathrm{B}}

\newcommand{\up}{\uparrow}
\newcommand{\down}{\downarrow}

\firstpage{1} 
\pubvolume{1}
\issuenum{1}
\articlenumber{0}
\pubyear{2026}
\copyrightyear{2026}
\datereceived{ } 
\daterevised{ } 
\dateaccepted{ } 
\datepublished{ } 

\Title{Quark Saturation in Magnetized Pure Neutron Matter}

\Author{ Toru Kojo $^{1,2\ddagger}$}

\AuthorNames{Firstname Lastname, Firstname Lastname and Firstname Lastname}

\address{%
$^{1}$ \quad Theory Center, IPNS, High Energy Accelerator Research Organization (KEK), 1-1 Oho, Tsukuba, Ibaraki, 305-0801, Japan\\
$^{2}$ \quad Graduate Institute for Advanced Studies, SOKENDAI, 1-1 Oho, Tsukuba, Ibaraki, 305-0801, Japan 
}

\corres{Correspondence: torukojo@post.kek.jp}

\abstract{
We study how magnetic fields affect the onset of quark saturation, which
limits the domain of applicability of a purely baryonic description.
We consider matter composed of electrically neutral $ddu$ baryons,
including neutron- and $\Delta^0$-like states.
Using a simple constituent-quark model, we compute the baryon wave
functions and spectra in magnetic fields.
The magnetic field polarizes the spin-flavor structure, 
compresses the internal wave function in the transverse direction, 
and strongly suppresses the transverse kinetic energy.
From the baryon wave functions, we construct the one-body reduced
density matrix and derive the corresponding single-quark distributions
in longitudinal momentum and Landau-level space.
We then use these distributions to evaluate quark occupation
probabilities in an ideal baryon gas and determine the onset of quark saturation.
At finite magnetic field in the $+z$-direction, the $d_\downarrow$ quark in the lowest
Landau level reaches saturation first.
The resulting saturation density depends nonmonotonically on the magnetic-field strength: 
it decreases at weak fields and increases at stronger fields.
This behavior reflects the competition between spin-flavor polarization, 
which promotes saturation, and transverse spatial squeezing, 
which delays saturation by broadening of the quark transverse momentum distribution.
}

\begin{document}

\section{Introduction}

At low densities, hadrons as composite particles are useful effective degrees of freedom
for describing many-body systems governed by quantum chromodynamics (QCD).
As the density increases, however, their quark and gluon substructure
becomes increasingly relevant to the interactions among hadrons and
eventually to the bulk properties of the system.
The change from a hadronic description to one in terms of quarks and gluons need not be sharp,
since the same underlying quark and gluon dynamics is responsible both for the internal
structure of hadrons and for their interactions \cite{Fukushima:2020cmk}.
At vanishing baryon density, lattice QCD with physical quark masses has established that the
finite-temperature transition is an analytic crossover
\cite{Aoki:2006we,Bazavov:2011nk}.
At low temperatures and high baryon densities, smooth hadron--quark crossover or continuity
has also been explored theoretically \cite{Schafer:1998ef,McLerran:2007qj} 
and phenomenologically in neutron-star matter
\cite{Masuda:2012kf,Masuda:2012ed,Baym:2017whm,Baym:2019iky,Kojo:2021wax}.

Recently, several approaches have incorporated the composite quark structure of baryons 
and quark-level Pauli constraints into descriptions of dense baryonic matter \cite{Jeong:2020QY,Duarte:2020EVQ,Duarte:2021QY,
Blaschke:2020QPB,Koch:2023Baryquark,Xia:2024NJL}.
In a class of such approaches, these constraints are formulated directly in terms of quark occupation probabilities. 
As density increases, low-momentum quark states eventually reach unit occupation, 
a phenomenon referred to here as {\it quark saturation} \cite{Kojo:2021PRD,Fujimoto:2024PRL,Fujimoto:2024doc,Bluhm:2025PRC,Kojo:2025vcq,Bluhm:2026stat}.
Beyond quark saturation, a quark Fermi sea begins to develop.
Somewhat surprisingly, quark-saturation effects can become important already around
$2$--$3 n_0$, where $n_0=0.16 \,{\rm fm}^{-3}$ is the nuclear saturation density.
This density is substantially lower than the naive geometric scale at which baryons with radii
of order $0.5\,{\rm fm}$ would strongly overlap.\footnote{
Similar observations have been obtained in model calculations for two-color and isospin QCD 
where many composite bosons of the size $\sim \lqcd^{-1}$ 
are generated by chemical potentials \cite{Chiba:2023ftg,Kojo:2024sca,Kojo:2021hqh}.
}
The detailed estimate of the onset of quark saturation, however, is sensitive to the size
and internal wave function of the baryons.

In this work, we study electrically neutral $ddu$ baryonic matter in uniform and static
magnetic fields with $|eB|\sim\Lambda_{\rm QCD}^2 \sim (200-300\,{\rm MeV})^2$,
corresponding to $B\sim 10^{19} \,{\rm G}$, or larger.
In this strong-field regime,
the spectra and internal structure of baryons 
are substantially reorganized \cite{Andreichikov:2014PRD,Taya:2014nha,Endrodi:2019whh,He:2016,Yakhshiev:2019,Chen:2023,Amari:2025,Coppola:2020qbh,Dominguez:2020sdf}.
For comparison, the surface dipole fields of magnetars are typically of order
$10^{14}$--$10^{15}\,{\rm G}$ \cite{Thompson:1995gw,Harding:2006qn,Kaspi:2017fwg}, 
while global equilibrium considerations suggest an upper bound 
on the average internal magnetic field of order $10^{18}\,\mathrm G$ \cite{Cardall:2000bs,Broderick:2001qw}
(although local fields may be substantially stronger than the large-scale field \cite{Bucciantini:2014mta}).
Microscopically, possible spontaneous magnetization in quark matter
has also been discussed, 
with the field strength reaching up to $\sim 10^{16}\,{\rm G}$ in some model studies
\cite{Tatsumi:2000Ferromag,Yoshiike:2015Mag,Tsue:2015Mag}.
Thus, even allowing for such exotic mechanisms, the field strengths considered here
are substantially larger than conventional estimates for magnetars,
and hence the direct phenomenological applications of our results may be limited.
We believe, however, that this study is valuable because
our setup poses conceptually interesting questions about
the interplay between hadronic structure and quark saturation,
and also provides a useful setting for exploring 
how dense matter evolves toward the strong-field regime, 
which may in turn offer qualitative insight into the behavior at more realistic field strengths.

We use a simple non-relativistic constituent-quark model 
\cite{DeRujula:1975qlm,Isgur:1979be}
to describe individual baryons and construct an ideal baryon gas in magnetic fields.
The applicability of this non-relativistic framework to low-lying states in strong magnetic fields,
as well as the approximation of a $B$-independent constituent-quark mass \cite{Kojo:2012js},
was discussed in detail in Ref.~\cite{Kojo:2026}.
Fields of $|eB| \sim\Lambda_{\rm QCD}^2$ substantially modify the baryon
energy spectra, including their dependence on the baryon pseudomomentum,
as well as the spin-flavor structure and spatial quark wave functions \cite{Andreichikov:2014PRD}.
The resulting baryonic Fermi sea becomes highly anisotropic.

The onset of quark saturation is estimated through the following simple sum rule.
At $B=0$, the quark occupation probability in baryonic matter is schematically obtained by
convoluting the baryon distribution $f_b$ with the quark distribution
$\phi_{q/b}$ inside a baryon ($b=n_\up, n_\down, \Delta^0_\up, \cdots$ and $q=u_\up, u_\down, d_\up, \cdots$)
\begin{align}
f_q (\bm{p})
= \sum_{b}
N_q^b 
\int_{\bm{K}}
f_b (\bm{K})\,
\phi_{q/b}(\bm{p};\bm{K}) \,.
\end{align}
Here $N_q^b$ specifies the quark spin-flavor content in a given baryon $b$.
The baryon distribution $f_b$ should in general be determined from the many-body dynamics of baryons.
In this work, we generalize this relation to finite $B$.
We first formulate the internal quark distribution using the one-body
reduced density matrix constructed from the single baryon wave function, 
and then construct the corresponding quark occupation probabilities in the Landau-level basis.
For simplicity, we restrict ourselves to an ideal baryon gas.
Within this framework, one can determine the density at which quark
saturation occurs and the purely baryonic description reaches its limit of applicability.

This paper is structured as follows.
In Sec.~\ref{sec:neutral_baryons} we discuss the properties of single baryons within a non-relativistic constituent quark model.
In Sec.~\ref{sec:quark_dist} we construct a density matrix from a baryon wave function and derive one-body reduced density matrix for a quark.
We then derive a quark distribution function for a given baryon.
In Sec.~\ref{sec:dense} we compute the quark distribution in dense baryonic matter and estimate the density at which quark saturation occurs.
Sec.~\ref{sec:summary} is devoted to summary.

\section{Neutral baryons in magnetic fields} \label{sec:neutral_baryons}

We first clarify the relation between particle labels and flavor labels.
Since the color wave function of a color-singlet baryon is totally antisymmetric, 
the remaining space-spin-flavor wave function is totally symmetric.
For the $ddu$ sector, we introduce a reference component,
\beq
|\Psi_{\rm ref}^{ddu}\ra
= \sum_{s_1,s_2,s_3}
\int_{\br_1,\br_2,\br_3}
\psi_{\rm ref}^{ddu} ( \br_1, \br_2, \br_3)
| d, {\bm \xi}_1; d, {\bm \xi}_2; u, {\bm \xi}_3 \ra \,,
~~~~~~
{\bm \xi}_j = (\br_j, s_j^z)
\eeq
%
and construct the physical state as
\beq
|\Psi^{ddu}\ra
=
\calN \calS_3 |\Psi_{\rm ref}^{ddu}\ra ,
\qquad
\calS_3 = \frac{1}{\, 3! \,}\sum_{\calP \in S_3} P \,.
\eeq
where $\calN$ is the normalization constant and $\calP$ is the permutation operator.
Thus, the assignment $(d_1,d_2,u_3)$ refers only to the reference
component; in the fully symmetrized state, flavor is not tied to a fixed particle slot.
Accordingly, the Hamiltonian should be regarded as an operator acting
also in flavor space.  We introduce the flavor projectors
\beq
\calP_f^{(j)}
= |f\ra_j\,{}_j\la f| ,
\qquad
f=u,d ,
\eeq
and write the one-body Hamiltonian associated with particle slot $j$ as
\beq
\hat h_j
=
\sum_{f=u,d}
\calP_f^{(j)} h_f(j),
\eeq
where [$\bm{\Pi}_f (j) \equiv \bp_j - e_f \bm{A} (\br_j)$]
\beq
h_f(j)
= m_f
+ \frac{\, \bm{\Pi}_f^2 (j) \,}{\, 2m_f \,}
- \frac{\, e_f \,}{\, 2m_f \,}
{\bm \sigma}_j \cdot{\bm B} \,,
\eeq
where 
we set $g=2$ for the constituent-quark magnetic moment, as in the standard non-relativistic constituent-quark model \cite{DeRujula:1975qlm,Isgur:1979be}. 
The corresponding neutron magnetic moment will be used below to fix the constituent-quark mass.
For instance, acting on the flavor state $|ddu\ra$, we obtain
\beq
\hat h_1 |ddu\ra = h_d (1) |ddu\ra,\qquad
\hat h_2 |ddu\ra = h_d (2) |ddu\ra,\qquad
\hat h_3 |ddu\ra = h_u (3) |ddu\ra .
\eeq
Since the full one-body Hamiltonian $\sum_{j=1}^3 \hat h_j$
is symmetric under permutations of the particle slots $j=1,2,3$,
the eigenvalue equations for $|ddu\ra$, $|dud\ra$, and $|udd\ra$
are related simply by relabeling the particle indices $j$.
Hence, the eigenvalues are unchanged under such permutations.
For the wave functions, once the wave function is obtained for a specific flavor ordering,
those for the other orderings follow by the corresponding permutations of the particle indices $j$.

The full Hamiltonian is then written as
\beq
\hat H
=
\sum_{j=1}^3 \hat h_j
+
V_{\rm conf}
+
V_{\rm short}.
\eeq
For the confining potential, we consider the sum of the harmonic oscillators
\beq
V_{\rm conf} = \alpha \sum_{i>j} \br_{ij}^2 + C_0 \,,~~~~~~~\alpha \sim \lqcd^3 \,.
\eeq
The constant $C_0$ represents a $B$-independent contribution to the
baryon energy and does not affect the wave functions.
We treat the confining potential at leading order, while the short-range potential
\beq
V_{\rm short} = \sum_{i>j} \big( v^E_{ij} + v_{ij}^M \bm{\sigma}_i \cdot \bm{\sigma}_j \big) \,,
\eeq
is regarded as a perturbation.
The explicit forms of the color-electric and magnetic interactions are given in Sec.~\ref{sec:baryon_dispersion}.

The interaction terms considered below are also symmetric under
permutations of particle slots $j=1,2,3$.  The full Hamiltonian therefore satisfies
\beq
P\hat H P^{-1}=\hat H,
\qquad P\in S_3 .
\eeq
For practical calculations, we may work with the reference
$d_1d_2u_3$ component.  Acting on this component, the flavor
projectors simply select
\beq
\hat h_1+\hat h_2+\hat h_3
\;\longrightarrow\;
h_d(1) + h_d(2) + h_u(3) \,.
\eeq
Therefore, within the reference component we may use
\beq
e_1=e_2=e_d,
\qquad
e_3=e_u,
\qquad
m_1=m_2=m_d,
\qquad
m_3=m_u ,
\eeq
while the full physical state is obtained by symmetrization.

\subsection{ Pseudomomentum }

In a uniform magnetic field, the ordinary canonical momentum is not the most convenient quantity for characterizing translational motion. 
To separate the collective motion from the internal dynamics, it is useful to introduce the pseudomomentum \cite{Johnson:1949,Avron:1978,Johnson:1983}
\beq
\bm{\calK}_j = \bm{\Pi}_j + e_j \bm{B} \times \bm{r}_j \,.
\eeq
The commutation rules among $\Pi$ and $\calK$ are
\beq
\big[ \Pi_i^x, \Pi_j^y \big] = \rmi e_i  \delta_{ij} B^z = - \big[ \calK_i^x, \calK_j^y \big] \,,
\eeq
and all the other commutators are zero; in particular, $\big[ \vec{\Pi}_i, \vec{\calK}_j \big] = 0$.

For a translationally invariant Hamiltonian, the total pseudomomentum $\bm{\calK}_R \equiv \sum_j \bm{\calK}_j$ commutes with the Hamiltonian,
$ [\bm{\calK}_R, H ] =0 $.
Meanwhile, in general the $x$,$y$ components of $\bm{\calK}_R$ do not commute with each other,
\beq
\big[ \calK_R^x, \calK_R^y \big]
= - \rmi B \sum_j e_j 
= - \rmi e_R B \,,~~~~~~~ e_R \equiv \sum_i e_i \,.
\eeq
However, for charge neutral states as we discuss in this paper,
this commutation relation vanishes.
Hence, the eigenstates of the Hamiltonian can be labelled by $(E,\bm{K}_R)$.

For a three-particle system, we choose Jacobi coordinates
\beq
\bm{R} = \frac{\, \sum_j m_j \br_j\,}{ M } \,,~~~~~ \bm{\rho} = \br_2 - \br_1 \,,~~~~~~ \bm{\lambda} = \br_3 - \frac{\, m_1 \br_1 + m_2 \br_2 \,}{ m_{12} } \,,
\eeq
where $m_{12} = m_1 + m_2$ and $M = m_1 + m_2 + m_3$.
Then we note
\beq
\sum_j e_j \br_j
= \bm{R} \sum_j e_j + \frac{\, e_2 m_1 - e_1 m_2 \,}{ m_{12} } \bm{\rho} 
+ \frac{\,  e_3 m_{12} - (e_1 + e_2 ) m_3 \,}{ M } \bm{\lambda} 
\equiv 
e_R \bm{R} + e_\rho \bm{\rho} + e_\lambda \bm{\lambda} 
\,.
\eeq
For neutrons and $\Delta_0$, taking the particle slots to be ordered as $ddu$, we set
$q \equiv e/3$, $e_d = - q$, and $e_u = 2 q$,
and $m_1=m_2=m_3=m$,
for which $e_R= 0 = e_\rho$ and $e_\lambda=2e/3=2q$.

Now, choosing the symmetric gauge, $\bm{A} = \bm{B} \times \br/2$,
one can write ($\bm{B}_q \equiv q \bm{B}$)
\beq
\bm{\calK}_R 
= \bm{P}_R + \sum_j  \frac{\, e_j \bm{B} \times \bm{r}_j \,}{2} 
= \bm{P}_R + \frac{\, e_\lambda \bm{B} \times \bm{\lambda} \,}{2} 
= \bm{P}_R + \bm{B}_q \times \bm{\lambda}  
\,.
\eeq
We consider the eigenfunction of the eigenvalue problem $ \bm{\calK}_R \Phi_{\bm{K}} = \bm{K} \Phi_{\bm{K}} $
whose solution takes the form
\beq
 \Phi_{\bm{K}} (\bm{R}, \bm{\rho}, \bm{\lambda})
  = \rme^{ \rmi \Theta (\bm{R},\bm{\lambda}) } \varphi_{\bm{K}} (\bm{\rho}, \bm{\lambda})
  \,,~~~~~~~~
\Theta (\bm{R},\bm{\lambda}) 
= \bm{R} \cdot \big( \bm{K} - \bm{B}_q \times \bm{\lambda}  \big)
\label{eq:def_Phi}
\,.
\eeq
The unitary transformation of the Hamiltonian, $H \rightarrow H' = \rme^{-\rmi \Theta} H \rme^{\rmi \Theta}$,
factorizes the center-of-mass coordinate, leaving the pseudomomentum \(\mathbf K\) as a parameter in the internal Hamiltonian.
The resulting Hamiltonian takes the form $H' = H_0' + V_{\rm short}$, 
where
\beq
H'_0 
= M + \frac{\, \bm{K}^2  \,}{\, 2M \,} + H_0^z + H_0^\perp + H_{\rm Zeeman}
 \,.
\eeq
Here the orbital terms in the $z$-direction are
\beq
H_0^z
= \frac{\, (p_\rho^z)^2  \,}{\, m \,}
+ \frac{\, 3\alpha \,}{2} \rho_z^2
+ \frac{\, 3(p_\lambda^z )^2  \,}{\, 4 m \,}
+ 2 \alpha \lambda_z^2
\,.
\eeq
The orbital terms in the transverse direction are
\beq
H_{0 \rho}^\perp
= \frac{\, (\bp_\rho^\perp)^2  \,}{\, m \,}
+ \bigg( \frac{\, B_q^2 \,}{16m} + \frac{\, 3\alpha \,}{2}  \bigg) \rho_\perp^2
+ \frac{\, 1 \,}{\, 2 m \,}  \bm{B}_q \cdot \bm{L}_\rho 
\,,
\eeq
for $\rho$ modes,
and, for $\lambda$ modes,
\beq
H_{0 \lambda}^\perp
=
\frac{\, 3  (\bp_\lambda^\perp -\bp_0^\perp)^2  \,}{\, 4 m \,}
+ \bigg(  \frac{\, 3 B_q^2 \,}{4m} + 2\alpha  \bigg) ( \bm{\lambda}_\perp - \bm{\lambda}^0_\perp)^2
- \frac{\, \bm{B}_q \cdot \tilde{\bm{L}}_\lambda  \,}{\, 2 m \,}  
- \frac{\, B_q^2 \bK_\perp^2 \,}{\, 6m (B_q^2 + 3 m \alpha) \,}
\,,
\label{eq:H0_lambda}
\eeq
where $\tilde{\bm{L}}_\lambda = (\bm{\lambda}_\perp -\bm{\lambda}^0_\perp) 
\times (\bp_\lambda^\perp -\bp_0^\perp)$ and $\bm{p}_0^\perp$ and $\bm{\lambda}^0_\perp$ 
and the constant shifts are
\beq
\bm{\lambda}_\perp^0 = \frac{\, \bK_\perp \times \bm{B}_q \,}{\, 2 (B_q^2 + 3 m \alpha) \,}
\,,~~~~~~~~~
\bp_0^\perp  
= \frac{\, \bm{B}_q \times \bm{\lambda}_\perp^0 \,}{\, 3 \,} 
=
\frac{\, B_q^2 \,}{\, 6 (B_q^2 + 3 m \alpha) \,} \bm{K}_\perp
\,.
\eeq
These shifts generate a quadratic term proportional to $\bm{K}_\perp^2$ in Eq.~\eqref{eq:H0_lambda}, which partially cancels $\bm{K}_\perp^2/2M$.
The Zeeman terms are
\beq
H_{\rm Zeeman}
= - \frac{\, \bm{B}_q \,}{\, 2 m \,}   \cdot \big[ 2 \bm{\sigma}_3 - ( \bm{\sigma}_1 + \bm{\sigma}_2) \big]
\,.
\eeq
Below we use the notations
\beq
B_\rho^2 \equiv B_q^2 + 24 m \alpha \,,~~~~~~~~~
B_\lambda^2 \equiv B_q^2 + 8 m \alpha/3 \,.
\eeq
We define $B_\rho$ and $B_\lambda$ to be positive, $B_\rho \equiv |B_\rho|$ and $B_\lambda \equiv |B_\lambda|$.
 
\subsection{ Baryon dispersion relations} \label{sec:baryon_dispersion}

Diagonalizing the unperturbed Hamiltonian $H_0$
and evaluating the short-range interaction to first order in perturbation theory,
the Hamiltonian in a given orbital state takes the form
\begin{align}
\hspace{-0.5cm}
E_{\bm{n},\{s\}} (\bm{K}) 
& = M
+ \frac{\, K_z^2  \,}{\, 2M \,} 
+ \frac{\, 2\alpha K_\perp^2 \,}{\, 4(B_q^2 + 3 m\alpha ) \,}
+ \sqrt{\frac{\, 6\alpha \,}{m} } \big( n_\rho^z + n_\lambda^z + 1 \big) 
\notag \\
&
~~+ \frac{\, B_\rho \,}{\, 2m \,} \bigg( 2n_\rho^\perp + |L_\rho^z| + \frac{\, B_q \,}{\, B_\rho \,} L_\rho^z + 1 \bigg)
+ \frac{\, B_\lambda \,}{\, 2m \,} \bigg( 6n_\lambda^\perp + 3 | L_\lambda^z | - \frac{\, B_q \,}{\, B_\lambda \,} L_\lambda^z + 3 \bigg)
\notag \\
& 
~~+ H_{\rm Zeeman}
+ \sum_{i>j} \big( \la v_{ij}^E \ra_{\bm{n}}^{\bK}  + \la v_{ij}^M \ra_{\bm{n}}^{\bK}  \bm{\sigma}_i \cdot \bm{\sigma}_j \big) 
 \,,
\end{align}
where $n_i^z$ and $n_i^\perp$ $(i=\rho, \lambda)$ are radial excitations 
while $L_i^z$ are azimuthal angular-momentum around the $z$-axis.
Below we focus on the states with minimal orbital excitations and set these quantum numbers to zero,
\beq
\hspace{-0.5cm}
E_{0,\{s\}} (\bm{K}) 
= E_{0} (\bm{K}) 
+ H_{\rm Zeeman}
+ \sum_{i>j} \la v_{ij}^M \ra_{0}^{\bK} \, \bm{\sigma}_i \cdot \bm{\sigma}_j 
\eeq
where the spin-independent part is
\beq
E_{0} (\bm{K}) 
 = M
+ \frac{\, K_z^2  \,}{\, 2M \,} 
+ \frac{\, 2\alpha K_\perp^2 \,}{\, 4(B_q^2 + 3 m\alpha ) \,}
+ \sqrt{\frac{\, 6\alpha \,}{m} } 
+ \frac{\, B_\rho + 3B_\lambda\,}{\, 2m \,} 
+ \sum_{i>j}  \la v_{ij}^E \ra_{0}^{\bK}
\,.
\eeq
We note that the \(K_\perp\) dependence arises from the potential terms: the confining potential generates the explicit quadratic term, 
while the short-range matrix elements also depend on \(K_\perp\) through the shifted internal wave function.

For later use, we give the explicit lowest-orbital wave function.
With the phase $\Theta$ defined in Eq.~\eqref{eq:def_Phi}, it is
\beq
\Phi_{\bm K}(\bm R,\bm\rho,\bm\lambda)
=
N_B e^{i\Theta(\bm R,\bm\lambda)}
e^{\rmi \bm p_0^\perp\cdot\bm\lambda_\perp}
\times
\exp\left[
- \frac{\, (\bm\lambda_\perp-\bm\lambda_\perp^0)^2 \,}{\, 2\beta_\lambda^\perp \,}
- \frac{\, \bm\rho_\perp^2 \,}{\, 2\beta_\rho^\perp \,}
- \frac{\, \lambda_z^2 \,}{\, 2\beta_\lambda^z \,}
- \frac{\, \rho_z^2 \,}{\, 2\beta_\rho^z \,}
\right]\,,
\label{eq:ground_wf}
\eeq
where
\beq
\beta_\lambda^\perp = \frac{1}{\, B_\lambda \,}\,,
\qquad
\beta_\rho^\perp = \frac{4}{\, B_\rho \,}\,,
\qquad
\beta_\lambda^z = \sqrt{ \frac{3}{\, 8m\alpha \,} }\,,
\qquad
\beta_\rho^z = \sqrt{\frac{2}{\, 3m\alpha \,} }\,,
\label{eq:ground_widths}
\eeq
and $N_B$ is the normalization constant.

\subsubsection{ Short-range interactions } \label{sec:short_range}

To evaluate the matrix elements, it is useful to introduce
\beq
\beta_{dd}^\perp=\beta_\rho^\perp \,,
\qquad
\beta_{dd}^z=\beta_\rho^z \,,
\label{eq:beta_dd_app}
\eeq
and
\beq
\beta_{ud}^\perp
=
\beta_\lambda^\perp + \frac{1}{\, 4 \,} \beta_\rho^\perp \,,
\qquad
\beta_{ud}^z
=
\beta_\lambda^z
+\frac{1}{\, 4 \,}\beta_\rho^z \,.
\label{eq:beta_ud_app}
\eeq
We characterize the typical scale of the short-distance interaction by
\beq
Q_{ij}^2 ~\equiv~  Q_0^2 +  (\beta_{ij}^z)^{-1} + (\beta_{ij}^\perp)^{-1} \,,
~~~~~~~~~ Q_0^2 \equiv 2m^2
\eeq
which reflects the typical inverse size of the corresponding quark pair.
For the running coupling, we use
\beq
\alpha_s(Q)
=
\frac{4\pi}{\, \beta_0\ln\left(Q^2/\Lambda_{\rm NP}^2\right) \,}
\,,
\qquad
\beta_0 = 11-\frac{\, 2N_f \,}{3}
\,.
\label{eq:alpha_running}
\eeq
We take $N_f=3$ and $\Lambda_{\rm NP}=0.25~{\rm GeV}$.
The renormalization scale $Q_{ij}$ is chosen to reflect the
typical inverse size of the corresponding quark pair $(ij)$.

For the short-range interactions, we use the color-electric and
color-magnetic potentials inspired by one-gluon exchange \cite{DeRujula:1975qlm,Isgur:1979be}.
The color-electric potential is
\beq
v^E_{ij}(r) = - \frac{2}{\, 3 \,}\,
\frac{\, \alpha_s(Q_{ij}) \,}{r}
\,.
\label{eq:VEM}
\eeq
The additional factor $1/2$ relative to the corresponding $q\bar q$
interaction reflects the color factor for a quark pair in the $\bar{\bm 3}$ channel.
For the color-magnetic interaction, 
we upgrade the standard non-relativistic form of the vertex as
$\alpha_s/m_i m_j \longrightarrow \alpha_s/E_iE_j$,
with $m_i = m_j = m$ and $\bp_i=\bp_j = \bq/2$. 
We then take the Fourier transform to get
[see Appendix.~\ref{app:matrix_elements} for details]
\beq
v^M_{ij}(r)
=
C_M \frac{\,  4 \alpha_s (Q_{ij}) \,}{\, 9 \,}
\frac{\, \rme^{-2mr} \,}{ r }
\,,
\label{eq:VS}
\eeq
where $C_M$ is a phenomenological parameter used for fine-tuning.

The short-range contributions are evaluated perturbatively as
\beq
\left\langle v^{E,M}_{ij}\right\rangle_0^{K_\perp}
=
\int \rmd^3 \br_1 \rmd^3 \br_2  \rmd^3 \br_3\,
\left|\Phi_{\bm K} (\bR, \bm{\rho}, \bm{\lambda} ) \right|^2
\, v^{E,M}_{ij}(r_{ij}) \,.
\label{eq:vij_expect}
\eeq
For the Gaussian ground-state wave function used here, these
matrix elements can be reduced analytically to closed expressions,
except for a one-dimensional integral in the displaced Coulomb matrix element. 
The explicit formulas are collected in Appendix~\ref{app:matrix_elements}.

Here we only note that, at small $K_\perp$,
the matrix element can be expanded in powers of $K_\perp^2$.
The $dd$ matrix elements are independent of $K_\perp$, 
while the $K_\perp$ dependence of the $ud$ matrix elements enters through
$(\lambda_\perp^0)^2\propto K_\perp^2/B^2$.  
Thus, for small $K_\perp$,
\beq
\left\langle v^{E,M}_{ud}\right\rangle_0^{K_\perp}
=
\left\langle v^{E,M}_{ud}\right\rangle_0^{K_\perp=0}
+
c^{E,M}_{ud}(B) K_\perp^2
+
{\cal O}(K_\perp^4) \,,
\label{eq:short_K_expansion}
\eeq
where the coefficient of the transverse term scales at large $B$ as
\beq
c^{E,M}_{ud}(B) \sim \alpha_s \lqcd/|B| \,.
\label{eq:short_K_expansion2}
\eeq
As in mesonic cases, the transverse kinetic energy term is suppressed by the factor $1/B$.

\subsection{Spin-flavor structure} \label{sec:spin-flavor}

For the lowest orbital state, the reference spatial wave function is
symmetric under the interchange of the two $d$ quarks
[$\Phi_K$ was defined in Eq.~\eqref{eq:def_Phi}],
\beq
\psi_{\rm ref}^{ddu}(\br_1,\br_2,\br_3) = \Phi_K(\bR,\brho,\blambda)
\,.
\eeq
which satisfies $\psi_{\rm ref}^{ddu}(\br_1,\br_2,\br_3) = \psi_{\rm ref}^{ddu}(\br_2,\br_1,\br_3)$.
For $S_z=-1/2$, we introduce the two normalized fully symmetric states
\beq
|\chi_0\ra
= \sqrt{3}\,\calS_3 |\tilde\chi_0\ra \,,
\qquad
|\chi_{-1}\ra
= \sqrt{3}\,\calS_3 |\tilde\chi_{-1}\ra \,,
\eeq
with
\begin{align}
|\tilde\chi_0\ra
&=
\int_{\br_1,\br_2,\br_3}
\psi_{\rm ref}^{ddu}
\frac{
|d_\up,\br_1;d_\down,\br_2;u_\down,\br_3\ra
+
|d_\down,\br_1;d_\up,\br_2;u_\down,\br_3\ra
}{\sqrt2} \,, 
\\
|\tilde\chi_{-1}\ra
&=
\int_{\br_1,\br_2,\br_3}
\psi_{\rm ref}^{ddu}
|d_\down,\br_1;d_\down,\br_2;u_\up,\br_3\ra \,.
\end{align}
At finite magnetic field, the spatial wave function is not in general
symmetric under permutations involving the $u$ quark.  Therefore,
the spin-flavor part does not factorize into a state of definite
isospin and total spin.

For later convenience, we introduce the spin basis with $S_z=-1/2$,
\begin{align}
|\Psi_{1/2}\ra
&\equiv
\frac{1}{\, \sqrt{3} \,}|\chi_0\ra
-
\sqrt{\frac{\, 2 \,}{\, 3 \,} } |\chi_{-1}\ra \,,
\\
|\Psi_{3/2}\ra
&\equiv
\sqrt{\frac{\,2\,}{\,3\,} } |\chi_0\ra
+
\frac{1}{\, \sqrt{3} \,}|\chi_{-1}\ra \,.
\end{align}
In the zero-field limit, where the usual spin-flavor classification is
recovered, these states continuously connect to the neutron and
$\Delta^0$, respectively.  At finite $B$, they are used simply as a
convenient basis for diagonalizing the spin-dependent Hamiltonian.

We now evaluate the spin-dependent part
\beq
H_{\rm spin}
=
H_{\rm Zeeman}
+
\sum_{i<j}
\la v^M_{ij}\ra_0^{\bK} \,
\bm\sigma_i\cdot\bm\sigma_j \,.
\eeq
In the spin basis introduced above, the color-magnetic interaction is diagonal,
\begin{align}
\la \Psi_{1/2} |
\sum_{i<j}
\la v^M_{ij}\ra_0^{\bK} \,
\bm{\sigma}_i\cdot\bm{\sigma}_j
| \Psi_{1/2} \ra
&=
\la v^M_{dd}\ra_0^{\bK}
-
4\la v^M_{ud}\ra_0^{\bK}
\equiv
V^M_{1/2} \,,
\\
\la \Psi_{3/2} |
\sum_{i<j}
\la v^M_{ij}\ra_0^{\bK} \,
\bm{\sigma}_i\cdot\bm{\sigma}_j
| \Psi_{3/2} \ra
&=
\la v^M_{dd}\ra_0^{\bK}
+
2\la v^M_{ud}\ra_0^{\bK}
\equiv
V^M_{3/2} \,,
\end{align}
while the off-diagonal matrix element vanishes,
\beq
\la \Psi_{1/2} |
\sum_{i<j} \la v^M_{ij}\ra_0^{\bK} \,
\bm{\sigma}_i\cdot\bm{\sigma}_j
| \Psi_{3/2} \ra = 0 \,.
\eeq
The Zeeman interaction, on the other hand, mixes the two basis states.
For $S_z=-1/2$, its matrix elements are
\begin{align}
\la \Psi_{1/2} | H_{\rm Zeeman} | \Psi_{1/2} \ra
& = - \frac{\, B_q \,}{m} \,,
\\
\la \Psi_{3/2} | H_{\rm Zeeman} | \Psi_{3/2} \ra
& = 0 \,,
\\
\la \Psi_{1/2} | H_{\rm Zeeman} | \Psi_{3/2} \ra
&=
\sqrt{2}\,\frac{\, B_q \,}{m} \,.
\end{align}
In the zero-field limit, these matrix elements are related to the
neutron and $\Delta^0$ magnetic moments and to the
$n$--$\Delta^0$ transition magnetic moment.
With $q=e/3$, the nonrelativistic quark model gives
\beq
\mu_n
= -\frac{e}{3m}\,,
~~~~~~~~~
\mu_{\Delta^0}
= 0\,,
~~~~~~~~~
\left|\mu_{n\Delta^0}\right|
= \frac{\, \sqrt{2}\,e \,}{\, 3m \,} \,.
\eeq
Here $\mu_n$ is defined in the conventional $S_z=+1/2$ convention.
The experimental value of the neutron magnetic moment motivates
our choice $m \simeq 0.33~{\rm GeV}$.

The spin-dependent Hamiltonian in the basis
$\{ |\Psi_{1/2}\ra, |\Psi_{3/2}\ra \}$ is therefore
\beq
H_{\rm spin}^{(\downarrow)}
=
\begin{pmatrix}
V^M_{1/2} - B_q/m
&
\sqrt{2}\, B_q/m
\\
\sqrt{2}\, B_q/m
&
V^M_{3/2}
\end{pmatrix} \,.
\eeq
Its eigenvalues are
\beq
V^\downarrow_{\pm}
=
\frac{\, V^M_{3/2} + V^M_{1/2} - B_q/m \,}{2}
\pm
\frac{1}{2}
\sqrt{\, \left( V^M_{3/2} - V^M_{1/2} + B_q/m \right)^2 + 8\left(\frac{\, B_q \,}{\, m \,} \right)^2 } \,.
\eeq
For the $S_z=+1/2$ sector, the sign of the Zeeman interaction is reversed.
The corresponding spin-dependent Hamiltonian is
\beq
H_{\rm spin}^{(\uparrow)}
=
\begin{pmatrix}
V^M_{1/2}+B_q/m
&
-\sqrt{2}\,B_q/m
\\
-\sqrt{2}\,B_q/m
&
V^M_{3/2}
\end{pmatrix} \,,
\eeq
with eigenvalues
\beq
V^\uparrow_{\pm}
=
\frac{V^M_{3/2}+V^M_{1/2}+B_q/m}{2}
\pm
\frac{1}{2}
\sqrt{
\left(
V^M_{3/2}-V^M_{1/2}-B_q/m
\right)^2
+
8\left(\frac{B_q}{m}\right)^2
}\,.
\eeq
Below we analyze the weak- and strong-field limits.

For weak magnetic fields, the eigenvalues can be expressed directly
in terms of the magnetic moments. Defining
\beq
\Delta_M(B)
\equiv
V^M_{3/2}(B)-V^M_{1/2}(B) \,,
\eeq
we find
\begin{align}
V_{-\downarrow}
&=
V^M_{1/2}
- |\mu_n|B
- \frac{\, |\mu_{n\Delta^0}|^2 \,}{\, \Delta_M(0) \,} B^2
+ O(B^3) \,,
\\
V_{-\uparrow}
&=
V^M_{1/2}
+
|\mu_n|B
- \frac{\, |\mu_{n\Delta^0}|^2 \,}{\, \Delta_M(0) \,} B^2
+ O(B^3) \,,
\\
V_{+\downarrow}
&=
V^M_{3/2}
+ \frac{\, |\mu_{n\Delta^0}|^2 \,}{\, \Delta_M(0) \,} B^2
+ O(B^3) \,,
\\
V_{+\uparrow}
&=
V^M_{3/2}
+ \frac{\, |\mu_{n\Delta^0}|^2 \,}{\, \Delta_M(0) \,} B^2
+ O(B^3) \,.
\end{align}
Thus, at weak magnetic fields, the neutron magnetic moment produces
the linear Zeeman splitting between the two neutron-like branches,
while the $n$--$\Delta^0$ transition magnetic moment produces the
quadratic level repulsion.  The latter lowers both neutron-like
branches and raises both $\Delta^0$-like branches.

At large $B$, the four eigenvalues behave as
\begin{align}
V^\downarrow_-
&~\to~
-\,2\frac{B_q}{m}
+
\frac{V^M_{3/2}+2V^M_{1/2}}{3} \,,
\\
V^\uparrow_-
&~\to~
-\frac{B_q}{m}
+
\frac{2V^M_{3/2}+V^M_{1/2}}{3} \,,
\\
V^\downarrow_+
&~\to~
\frac{B_q}{m}
+
\frac{2V^M_{3/2}+V^M_{1/2}}{3} \,,
\\
V^\uparrow_+
&~\to~
2\frac{B_q}{m}
+
\frac{V^M_{3/2}+2V^M_{1/2}}{3} \,.
\end{align}
The spin composition of this lowest branch also becomes particularly
simple.  For $B>0$,
\beq
|\widetilde n_{\downarrow}\ra
\longrightarrow
|\chi_{-1}\ra
\sim
|d_{\downarrow}d_{\downarrow}u_{\uparrow}\ra ,
\qquad
B\to\infty .
\eeq
Thus the strong magnetic field polarizes the spin-flavor composition
toward the configuration
$d_{\downarrow}d_{\downarrow}u_{\uparrow}$.
As will be discussed below, this polarization plays an important role
in determining which quark state first reaches saturation in baryonic matter.

Combining the spin-dependent eigenvalues with the spin-independent
orbital energy, the four branches are defined as
($s=\uparrow,\downarrow$)
\begin{align}
E^{\widetilde n}_{0s}(\bK)
\equiv
E_0(\bK)+V^s_- \,,
~~~~~~~~~~~~
E^{\widetilde\Delta^0}_{0s}(\bK)
\equiv
E_0(\bK)+V^s_+ \,.
\end{align}
Here $\widetilde n$ and $\widetilde\Delta^0$ denote the branches
adiabatically connected to the neutron and $\Delta^0$ states at $B=0$,
respectively.

At large $B$, the lowest branch behaves as
\beq
E^{\widetilde n}_{0\downarrow}(\bK)
\to
M
+
\frac{K_z^2}{2M}
+
\frac{\alpha K_\perp^2}{2B_q^2}
+
\sqrt{\frac{6\alpha}{m}}
+
\sum_{i>j}\la v^E_{ij}\ra_0^{\bK}
+
\frac{V^M_{3/2}+2V^M_{1/2}}{3} \,.
\label{eq:largeB_spectrum}
\eeq
The leading $O(B_q)$ contribution from the transverse zero-point energy
is canceled by the Zeeman energy in this branch.
The other three branches are separated from the lowest branch as
\begin{align}
E^{\widetilde n}_{0\uparrow}(\bK)
- E^{\widetilde n}_{0\downarrow}(\bK)
&~\to~
\frac{B_q}{m}
+ \frac{\Delta_M(B)}{3} \,,
\\
E^{\widetilde\Delta^0}_{0\downarrow}(\bK)
- E^{\widetilde n}_{0\downarrow}(\bK)
&~\to~
3\frac{B_q}{m}
+ \frac{\Delta_M(B)}{3} \,,
\\
E^{\widetilde\Delta^0}_{0\uparrow}(\bK)
- E^{\widetilde n}_{0\downarrow}(\bK)
&~\to~
4\frac{B_q}{m} \,.
\end{align}
Since $\Delta_M(B)>0$, the
$\widetilde n_\downarrow$ branch is the lowest of the four branches
in the strong-field regime.

The results above determine the orbital and spin-flavor structure
of the low-lying neutral baryon states in a magnetic field.
We next illustrate the resulting spectra for representative parameter sets.

\begin{table}[t]
\centering
\caption{
Parameter sets used for the numerical calculations.
The constituent-quark mass is fixed to $m=0.33~{\rm GeV}$.
For each set, the confinement parameter $\alpha$ is chosen to reproduce
the specified zero-field baryon-number rms radius $r_B(0)$, while
$C_M$ and $C_0$ are adjusted to reproduce
$M_N=0.939~{\rm GeV}$ and $M_\Delta=1.232~{\rm GeV}$ at $B=0$.
}
\label{tab:baryon_parameters}
\begin{tabular}{c|ccccc}
\hline
set
& $m$ [GeV]
& $r_B(0)$ [fm]
& $\alpha^{1/3}$ [GeV]
& $C_M$
& $C_0$ [GeV]
\\
\hline
A & 0.330 & 0.50 & 0.2305 & 2.2513 & -0.8286 \\
B & 0.330 & 0.60 & 0.1808 & 3.1641 & -0.4441 \\
C & 0.330 & 0.70 & 0.1472 & 4.3314 & -0.2241 \\
\hline
\end{tabular}
\end{table}

\subsection{Numerical illustration}
\label{sec:numerical_spectra}

\begin{figure}[tb]
\centering
\includegraphics[width=0.6\linewidth]{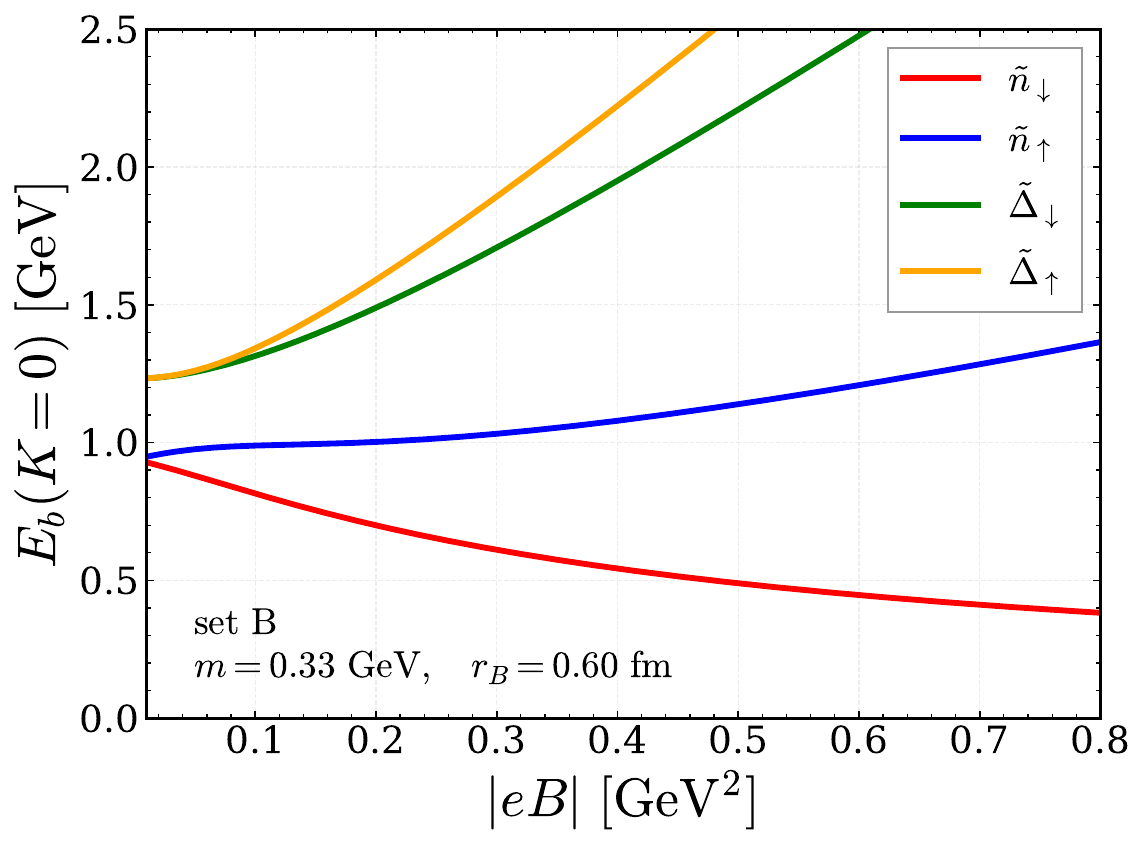}
\caption{
Magnetic-field dependence of the four low-lying neutral $ddu$ baryon
branches at $\bK=0$ for parameter set B.
The branches denoted by $\widetilde n$ and $\widetilde\Delta^0$ are
adiabatically connected to the neutron and $\Delta^0$ states at $B=0$,
respectively.
The magnetic field lifts the spin degeneracy, and the
$\widetilde n_\downarrow$ branch becomes the lowest-energy state as
$|eB|$ increases.
}
\label{fig:baryon_branches_setB}
\end{figure}
\begin{figure}[thb]
\centering
\includegraphics[width=0.495\linewidth]{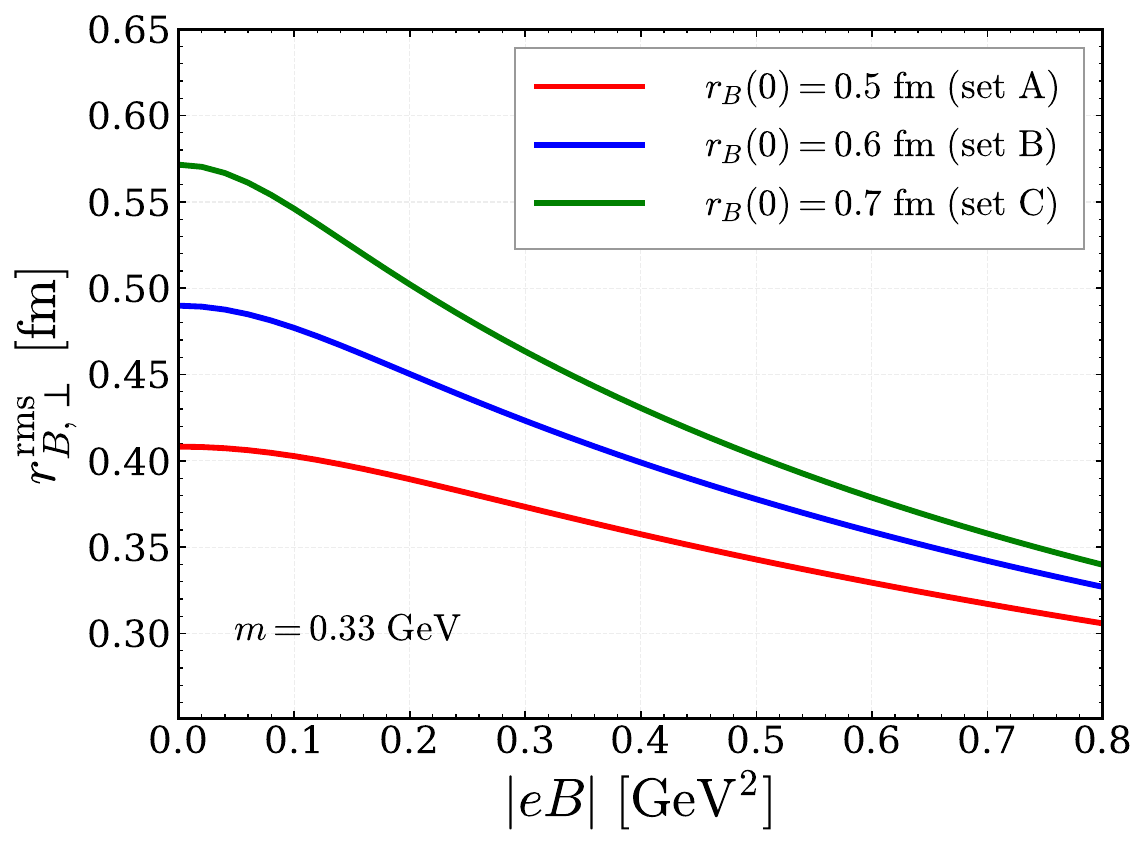}
\includegraphics[width=0.495\linewidth]{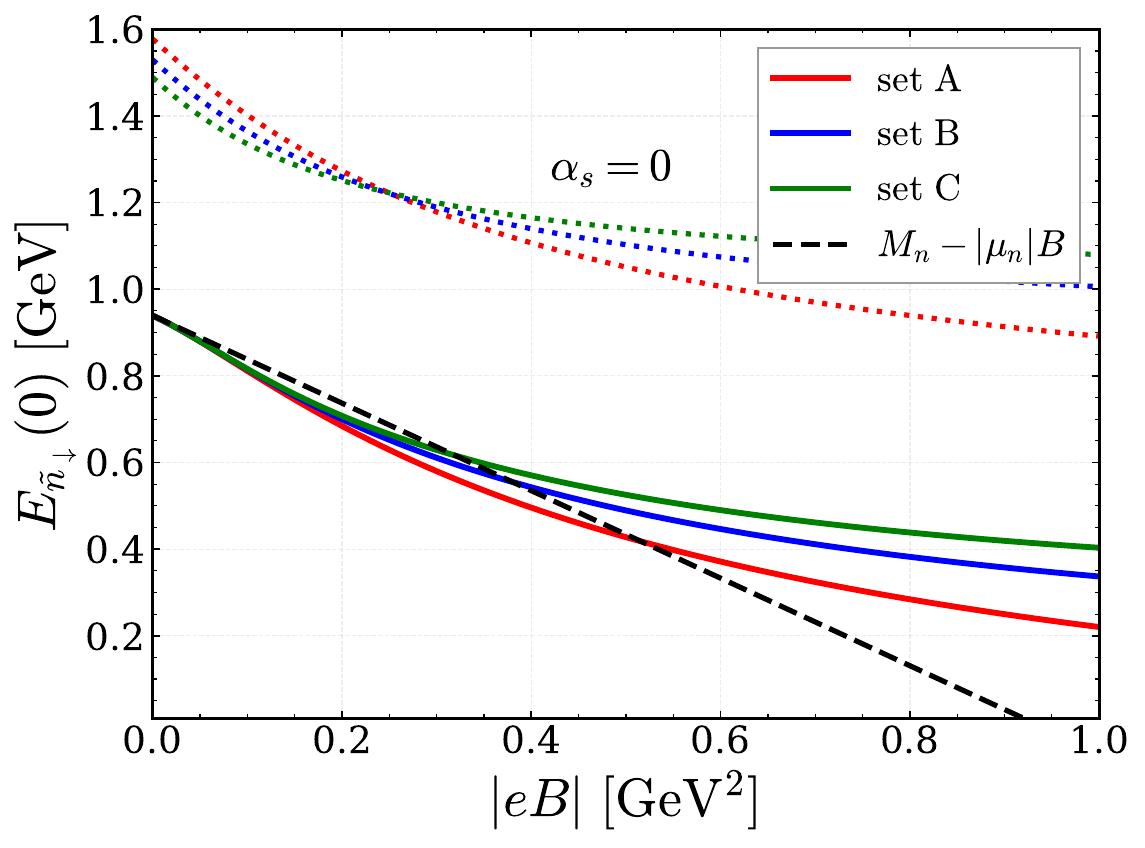}
\caption{
Magnetic-field dependence of the transverse size and ground-state
energy of the lowest $\widetilde n_\downarrow$ branch for parameter
sets A--C.
(Left panel) Transverse rms baryon radius,
$r_{B,\perp}^{\rm rms}$.
The magnetic field compresses the baryon in the transverse direction,
with the three curves corresponding to the different zero-field
baryon radii.
(Right panel) Ground-state energy
$E_{\widetilde n_\downarrow}(\bK=0)$.
Solid curves show the full results including the running short-range
coupling $\alpha_s(Q)$, while dotted curves show the corresponding
results with the short-range interaction switched off,
$\alpha_s=0$, without refitting the model parameters.
The black dashed line indicates the weak-field behavior
$M_n-|\mu_n|B$.
}
\label{fig:baryon_mass_B}
\end{figure}

For the numerical illustrations, we first fix the constituent-quark
mass using the neutron magnetic moment as a guide.
In the present nonrelativistic quark model,
\beq
\mu_n=-\frac{e}{3m}.
\eeq
The experimental value,
\beq
\mu_n=-1.91304276(45)\,\mu_N,
\qquad
\mu_N\equiv\frac{e}{2M_N},
\eeq
corresponds to
\beq
m\simeq 0.327~{\rm GeV}.
\eeq
We therefore fix $m=0.33~{\rm GeV}$ throughout the numerical analysis
and examine the sensitivity to the spatial size of the baryon.
We define
\beq
\la \br_B^2 \ra \equiv \frac{1}{\, 3 \,} \sum_{i=1}^3 \big\la (\br_i - \bR )^2 \big\ra
\eeq
and consider three values of the zero-field baryon radius,
\beq
r_B(0) \equiv \sqrt{ \la \br_B^2 \ra } = 0.50,\ 0.60,\ 0.70~{\rm fm},
\eeq
which define parameter sets A, B, and C, respectively.
For the Gaussian wave function for the ground state at $B=0$, the baryon radius is related
to the confinement parameter by
\beq
\alpha
=
\frac{1}{6m\,r_B^4(0)} \,.
\label{eq:alpha_from_rB}
\eeq
Thus, once $m$ and $r_B(0)$ are specified, the confinement parameter
$\alpha$ is fixed.
For each set, the confinement parameter $\alpha$ is determined from
the chosen value of $r_B(0)$, while $C_M$ and $C_0$ are adjusted to
reproduce
\beq
M_N=0.939~{\rm GeV},
\qquad
M_\Delta=1.232~{\rm GeV}
\eeq
at $B=0$.

Figure~\ref{fig:baryon_branches_setB} shows the magnetic-field dependence
of the four low-lying neutral $ddu$ baryon branches for the representative
parameter set B.
At $B=0$, the two lower and two upper branches reduce to the
spin-degenerate neutron and $\Delta^0$ states, respectively.
As the magnetic field increases, the spin degeneracy is lifted and the
$\widetilde n_\downarrow$ branch becomes the lowest-energy state.
Its relatively weak variation at strong fields reflects the cancellation
between the transverse orbital contribution and the Zeeman energy
discussed in Eq.~\eqref{eq:largeB_spectrum}.
The remaining branches are pushed upward relative to
$\widetilde n_\downarrow$, consistently with the strong-field behavior
derived above.

In the following, we focus mainly on the lowest
$\widetilde n_\downarrow$ branch, which provides the relevant
single-baryon state for the many-body analysis.

Figure~\ref{fig:baryon_mass_B} shows the transverse size 
$r_{B,\perp}^{\rm rms} \equiv \sqrt{ \la \br_\perp^2 \ra } $
and the ground-state energy of the lowest $\widetilde n_\downarrow$ branch
for parameter sets A--C.
As the magnetic field increases, the baryon is compressed in the
transverse direction, reflecting the magnetic localization of the
internal quark motion.
Although the quantitative behavior depends on the zero-field baryon
size, the same trend is observed for all three parameter sets.

\begin{figure}[tb]
\centering
\includegraphics[width=0.495\linewidth]{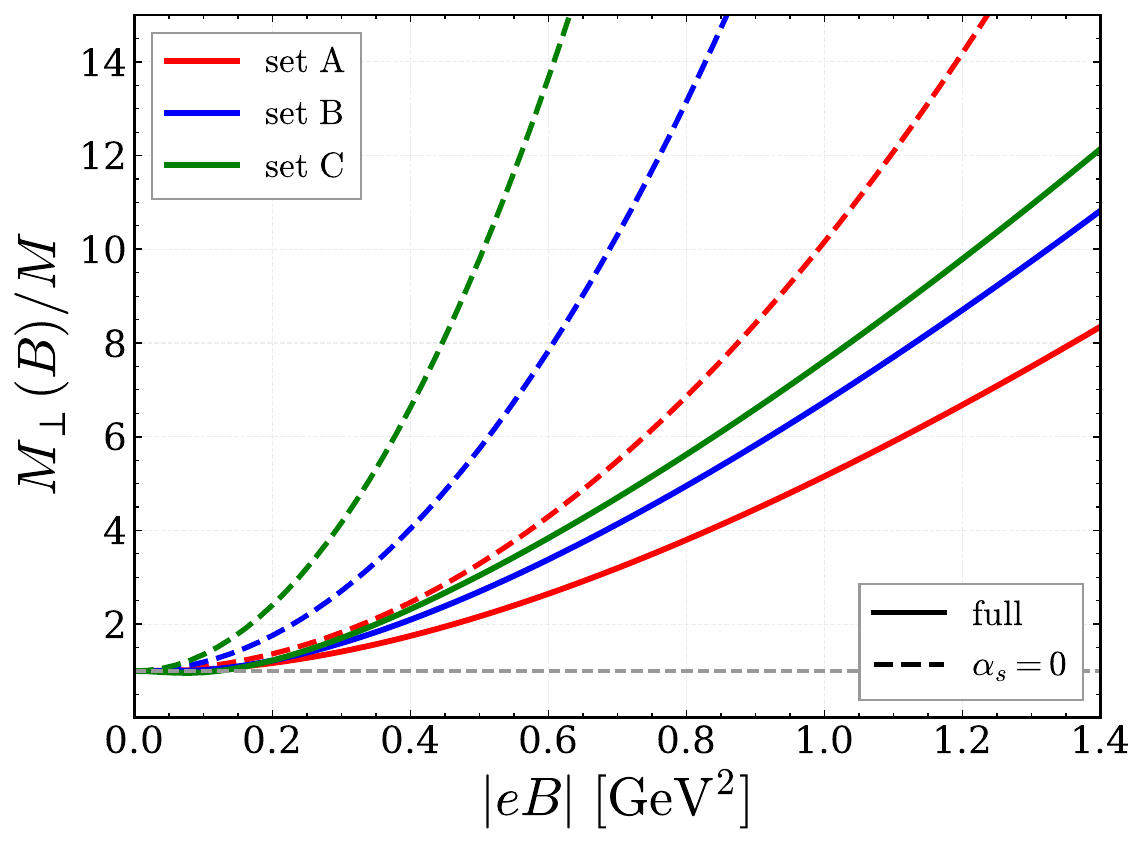}
\includegraphics[width=0.495\linewidth]{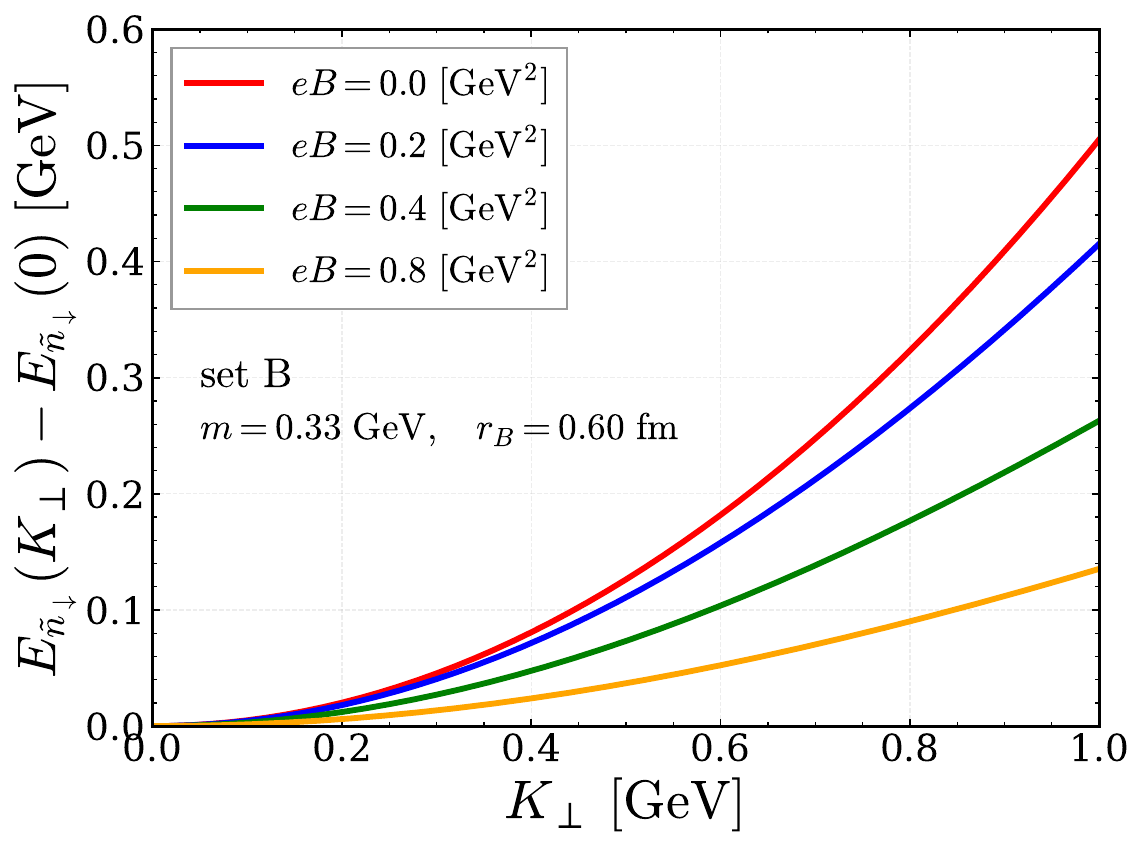}

\caption{ Transverse dispersion and effective mass of the lowest
$\widetilde n_\downarrow$ branch.
(Left panel) Transverse effective-mass ratio $M_\perp(B)/M$
for parameter sets A--C, extracted from the small-$K_\perp$
dispersion,
$E_{\widetilde n_\downarrow}(K_\perp,0)
-E_{\widetilde n_\downarrow}(0,0)
=
K_\perp^2/[2M_\perp(B)]+O(K_\perp^4)$.
Solid curves show the full results including the short-range
interaction, while dashed curves show the corresponding results
with $\alpha_s=0$.
The rapid increase of $M_\perp$ reflects the suppression of the
transverse baryon kinetic energy at strong magnetic fields.
(Right panel) Transverse excitation energy
$E_{\widetilde n_\downarrow}(K_\perp,0)
-E_{\widetilde n_\downarrow}(0,0)$
for parameter set B at several magnetic fields.
With increasing $|eB|$, the transverse dispersion becomes
progressively flatter.
}
\label{fig:Mperp_setB}
\end{figure}

The right panel of Fig.~\ref{fig:baryon_mass_B}
shows the corresponding ground-state energy.
Its sizable decrease with increasing $B$ originates from the
cancellation between the transverse orbital energy and the Zeeman
energy in the lowest branch.
This mechanism is particularly transparent when the short-range
interaction is switched off.
The full results including the running short-range coupling
$\alpha_s(Q)$ exhibit the same qualitative behavior, with additional
modifications associated with the magnetic-field dependence of the
short-range interactions.
As shown explicitly in Eq.~\eqref{eq:largeB_spectrum}, the leading
$O(B_q)$ contribution from the transverse zero-point motion is canceled
by the Zeeman term at large $B$.
Consequently, a substantial part of the transverse zero-point energy
present at $B=0$ is effectively removed from the lowest branch.

This effect is more pronounced for baryons than for mesons because a
three-quark system contains two independent internal transverse modes,
associated with the Jacobi coordinates $\brho$ and $\blambda$, whereas
a meson has only one relative transverse mode.
For comparison, at $|eB|\simeq 1~{\rm GeV}^2$ the ground-state energy
of a neutral meson decreases by about $200$--$300~{\rm MeV}$ \cite{Kojo:2021,Kojo:2026}, 
while the present baryon calculation gives a reduction of about
$400$--$600~{\rm MeV}$.
The approximately factor-of-two difference is therefore qualitatively
consistent with the number of independent transverse internal modes.

To characterize the magnetic-field dependence of the transverse
baryon motion, we define the transverse effective mass $M_\perp(B)$
from the small-$K_\perp$ dispersion of the lowest
$\widetilde n_\downarrow$ branch,
\beq
E_{\widetilde n_\downarrow}(K_\perp,0)
-
E_{\widetilde n_\downarrow}(0,0)
=
\frac{K_\perp^2}{2M_\perp(B)}
+
O(K_\perp^4) \,.
\label{eq:Mperp_def}
\eeq
At $B=0$, rotational invariance restores the isotropic dispersion, giving $M_\perp (B=0) = M =3m$.

Figure~\ref{fig:Mperp_setB} shows the magnetic-field dependence of
$M_\perp$ together with the transverse dispersion of the lowest branch.
The left panel shows $M_\perp(B)/M$ for parameter sets A--C.
With increasing $|eB|$, $M_\perp$ grows rapidly for all three parameter
sets, indicating a strong suppression of the transverse baryon motion.
The same qualitative behavior is found with and without the
short-range interaction, although the latter modifies the quantitative
behavior of $M_\perp$.
The right panel shows the transverse excitation energy
\beq
E_{\widetilde n_\downarrow}(K_\perp,0)
-
E_{\widetilde n_\downarrow}(0,0)
\eeq
for parameter set B at several magnetic fields.
As the magnetic field increases, the dispersion becomes progressively
flatter, directly illustrating the increase of $M_\perp$ seen in the left panel.
In the strong-field limit and $\alpha_s=0$, 
the explicit orbital contribution in Eq.~\eqref{eq:largeB_spectrum}
is suppressed as $K_\perp^2/B_q^2$.
The short-range interactions generate additional $K_\perp$-dependent terms
that decrease more slowly with $B$, as discussed in Eq.~\eqref{eq:short_K_expansion} and \eqref{eq:short_K_expansion2}.
A similar suppression of the transverse kinetic energy and flattening
of the dispersion has been found in mesonic systems
~\cite{Fukushima:2012kc,Hattori:2015aki,Kojo:2021,Kojo:2026}.

\section{ Quark distribution in a baryon: Density matrix } \label{sec:quark_dist}

In this section, we characterize the single-quark distribution starting
from the one-body reduced density matrix. We first keep the finite-volume
normalization factors explicit, so that matrix elements in individual
momentum and Landau states can be interpreted as probabilities for
discrete one-particle states. We later define the corresponding
volume-independent distributions by factoring out these normalization
factors.

\subsection{One-body reduced density matrix}

We now construct the single-quark density matrix starting from the
fully symmetrized baryon state.
For the $S_z=-1/2$ sector, we write the baryon state as
\beq
|\Psi_K^{(ddu)\downarrow}\ra
=
c_0 |\chi_0;K\ra
+
c_{-1} |\chi_{-1};K\ra \,,
\qquad
|c_0|^2+|c_{-1}|^2=1 \,,
\eeq
where the orbital dependence of the states is specified by the
reference wave function $\psi_{\rm ref}^{ddu}$ obtained in the previous section.
The corresponding baryon density operator is
\beq
\hat\rho_B
=
|\Psi_K^{(ddu)\downarrow}\ra
\la\Psi_K^{(ddu)\downarrow}| \,.
\eeq
The reduced density matrix associated with particle slot $i$ is
\beq
\hat\rho_i
= {\rm Tr}_{j\neq i}\hat\rho_B \,,
\qquad
i= 1, 2, 3 \,.
\eeq
Because the full baryon state is totally symmetric under permutations
of particle slots, the three reduced density matrices are equivalent,
$\hat\rho_1 = \hat\rho_2 = \hat\rho_3 $.
We therefore define the number-normalized one-body density operator
\beq
\hat\Gamma
\equiv
\sum_{i=1}^3 \hat\rho_i \,,
~~~~~~~~~
{\rm Tr}\,\hat\Gamma=3 \,.
\eeq
To extract the distribution of a given flavor and spin, we introduce
the projector
\beq
\calP_{fs} = | f,s \ra \la f,s| \,,
~~~~~~~~~
f = u,d,\qquad s = \up \,,\down \,,
\eeq
and define the corresponding one-body kernel by
\beq
\Gamma_{fs}(\br,\br')
= \la f, s, \br | \hat\Gamma| f, s,\br'\ra \,.
\eeq
For the two basis states introduced in Sec.~2.3, the spin-flavor
components of the one-body density matrix are
\begin{align}
\Gamma_{u\downarrow}(\br,\br')
&=
|c_0|^2 R_u(\br,\br'),
\\
\Gamma_{u\uparrow}(\br,\br')
&=
|c_{-1}|^2 R_u(\br,\br'),
\\
\Gamma_{d\uparrow}(\br,\br')
&=
|c_0|^2 R_d(\br,\br'),
\\
\Gamma_{d\downarrow}(\br,\br')
&=
\left(
|c_0|^2+2|c_{-1}|^2
\right)
R_d(\br,\br').
\end{align}
The spatial kernels are
\begin{align}
R_u(\br,\br')
&=
\int_{\br_1,\br_2}
\psi_{\rm ref}^{ddu} (\br_1,\br_2,\br)
\psi_{\rm ref}^{ddu*} (\br_1,\br_2,\br'),
\\
R_d(\br,\br')
&=
\int_{\br_2,\br_3}
\psi_{\rm ref}^{ddu} (\br,\br_2,\br_3)
\psi_{\rm ref}^{ddu*} (\br',\br_2,\br_3),
\end{align}
where the equality of the two $d$-quark kernels follows from the
$1\leftrightarrow2$ symmetry of the reference orbital wave function.
The interference terms between $|\chi_0\ra$ and $|\chi_{-1}\ra$
vanish in the spin-diagonal one-body density matrix after tracing over
the other two quarks.

After summing over spin, the flavor-resolved one-body kernels are
\beq
\Gamma_u(\br,\br')
= R_u(\br,\br') \,,
\qquad
\Gamma_d(\br,\br')
= 2R_d(\br,\br') \,,
\eeq
and hence
\beq
{\rm Tr}\,\Gamma_u=1 \,,
\qquad
{\rm Tr}\,\Gamma_d=2 \,.
\eeq
For the analysis of the momentum distribution of a single quark, it is
convenient to introduce density matrices normalized to unity,
\beq
\rho_K^u(\br,\br')
\equiv
\Gamma_u(\br,\br')
= R_u(\br,\br') \,,
\eeq
and
\beq
\rho_K^d(\br,\br')
\equiv
\frac{1}{\, 2 \,}\Gamma_d(\br,\br')
= R_d(\br,\br') \,,
\eeq
so that
\beq
{\rm Tr}\,\rho_K^u
= {\rm Tr}\,\rho_K^d
= 1 \,.
\eeq
%

\subsection{Wave functions and single-quark momentum distributions}

Using the explicit reference orbital wave function Eq.~\eqref{eq:ground_wf}, we now evaluate the normalized single-quark density matrices $\rho_{\bm K}^i$.
For $\rho_{\bm K}^i$ ($i=u,d$), explicit calculations yield
\beq
\rho_{\bm{K}}^i ( \br; \br' ) 
= |\calN_i |^2\,
 \rme^{\rmi \Xi_{ {\bm K} }^{i} (\br_\perp,\br_\perp') } \, \rme^{ \rmi {\bm \kappa}^i_\perp \cdot {\bm \xi}_\perp - A_\perp^i {\bm \xi}_\perp^2 }
\otimes \rme^{ \rmi \kappa_z^i \xi_z- A^i_z \xi_z^2 }
= \rho_{ {\bm K}_\perp }^i \otimes \rho_{ K_z }^i
\,,
\eeq
where $\bm{\xi} = \bm{r} - \bm{r}' $ and $\calN_i = (L_z S_\perp)^{-1/2}$, with $L_z$ the size in $z$-direction and $S_\perp$ the transverse area, so that $\tr[\rho_{\bm{K}}^i ] = 1$. 
Accordingly, the transverse and longitudinal parts scale as $\rho_{ {\bm K}_\perp }^i \propto 1/S_\perp$ and  $\rho_{ K_z }^i \propto 1/L_z$, respectively.
The density matrix factorizes into the transverse and longitudinal parts.
Functions in the exponent are ($i=u,d$)
\begin{align}
&\Xi_{ {\bm K} }^i (\br_\perp,\br_\perp')
= - \frac{\, e_i {\bm B} \,}{2} \cdot ( {\bm r}_\perp \times {\bm r}_\perp' )
\,,
~~~~~~~~
{\bm \kappa}^i
= \bigg( \frac{\, m\alpha \,}{\, B_q^2 + 3m \alpha \,} {\bm K}_\perp \,,  \frac{\, K_z \,}{3} \bigg) 
\,,~~~~~~~
\notag \\
&
A_z^i = \sqrt{ \frac{\, m \alpha \,}{\, 6 \,} } 
\,,~~~~~~~
A_\perp^u
= \frac{1}{\, 4 \,} \bigg( B_\lambda + \frac{\, B_q^2 \,}{\, B_\lambda \,} \bigg)
\,,~~~~~~~
A_\perp^d
= \frac{1}{\, 16 \,} \bigg( B_\rho + B_\lambda + \frac{\, B_q^2 \,}{\, B_\rho \,} + \frac{\, B_q^2 \,}{\, B_\lambda \,} \bigg)
\,.
\end{align}
The density matrix is not diagonal for $(\br,\br')$.

The one-body density matrix contains the complete information on the
single-quark distribution. For an arbitrary normalized one-particle
state $|a\ra$, the diagonal matrix element
$\la a|\hat\rho_{\bm K}^i|a\ra$ gives the probability for the quark to occupy that state.
It is therefore useful first to diagonalize $\hat\rho_{\bm K}^i$,
\beq
\hat\rho_{\bK}^i
=
\sum_b F_{b;\bK}^i\, |b;\bK\ra\la b;\bK| \,,
\eeq
where $|b;\bK\ra$ denotes an eigenstate of the one-body density matrix.
Then,
\beq
\la a | \hat\rho_{\bm K}^i |a \ra
=
\sum_b F_{b;\bK}^i\,
|\la a|b;\bK\ra|^2 \,.
\eeq
Thus, the eigenstates $|b;\bK\ra$ form the natural one-particle basis,
and the corresponding eigenvalues $F_{b;\bK}^i$ give the occupation
probabilities in that basis.
The distribution in any other basis, with basis states denoted by
$|a\ra$, is obtained by projecting $\hat\rho_{\bK}^i$ onto that basis.
We will subsequently express these probabilities in the momentum basis
along the magnetic field and in a fixed Landau basis in the transverse plane.


The natural-state label $b$ introduced above is resolved here into
transverse and longitudinal quantum numbers. Since
$\rho_{\bm K}^i$ factorizes into transverse and longitudinal parts,
the corresponding eigenvalue problems can be solved independently:
\begin{align}
 \int_{\br'_\perp }
 \rho_{{\bm K}_\perp}^i(\br_\perp,\br'_\perp)
 \psi^i_{{\bm n}_\perp;{\bm K}_\perp}(\br'_\perp)
&=
\frac{F^i_{{\bm n}_\perp;{\bm K}_\perp}}{S_\perp}
\psi^i_{{\bm n}_\perp;{\bm K}_\perp}(\br_\perp) \,,
\notag\\
 \int_{z'}
 \rho_{K_z}^i(z-z')
 \psi^i_{n_z;K_z}(z')
&=
\frac{F^i_{n_z;K_z}}{L_z}
\psi^i_{n_z;K_z}(z) \,.
\label{eq:def_f}
\end{align}
These eigenfunctions, $\psi^i_{{\bm n}_\perp;{\bm K}_\perp}$ and
$\psi^i_{n_z;K_z}$, are the coordinate-space representations of the
natural one-particle states $|b;\bK\ra$ introduced above.
For each flavor $i=u,d$, the transverse problem is formulated in the
Landau basis associated with the charge $e_i$, while the longitudinal
problem is diagonal in the momentum basis.
We discuss the two sectors in turn.

\subsubsection{ The $z$-component }

The $z$-component is relatively simple because
$\rho_{K_z}^i(z-z')$ depends only on the coordinate difference.
For such cases the natural one-particle eigenstates are the momentum
eigenstates,
\beq
\psi^i_{p_z;K_z}(z)=\rme^{\rmi p_z z} \,.
\eeq
Indeed, substituting this function into Eq.~\eqref{eq:def_f}, we obtain
\beq
 \int_{z'} \rho_{K_z}^i(z-z') \rme^{\rmi p_z z'}
 =
 \bigg[
 \int_{z'} \rho_{K_z}^i(z-z')
 \rme^{-\rmi p_z(z-z')}
 \bigg]
 \rme^{\rmi p_z z}
 =
 \frac{F^i_{p_z;K_z}}{L_z}
 \rme^{\rmi p_z z} \,.
\eeq
%
Thus, after factoring out the finite-volume normalization \(1/L_z\), the corresponding longitudinal distribution is
\beq
 F^i_{p_z;K_z}
 =
 \sqrt{\frac{\pi}{A_z^i}}\,
 \exp\bigg[
 -\frac{1}{4A_z^i}
 \bigg(p_z-\frac{K_z}{3}\bigg)^2
 \bigg]
 \equiv
 \varphi^i(p_z-K_z/3) \,.
\eeq
%

\subsubsection{The transverse component}

Solving the eigenvalue equation in Eq.~\eqref{eq:def_f} for the transverse direction
is more complicated than in the $z$ direction.
Unlike in the $z$ direction, ordinary translational invariance is lost,
and the momentum basis does not diagonalize the density matrix.
There is, however, magnetic translation symmetry \cite{Zak:1964a,Zak:1964b,Hongo:2026mag}, 
which allows us to identify a natural basis for the transverse motion.

In the current context, magnetic translation symmetry follows from
the conservation of the total pseudomomentum
$\hat{\bm{\calK}}_R$.
The full baryon wave function $\Phi_{\bK}$ and the corresponding
density matrix $\hat{\rho}_{\bK}^B$ satisfy
\beq
\rme^{\rmi \hat{ \bm{\calK} }_R \cdot \bm{a} } \Phi_{\bK} = \rme^{\rmi \bK \cdot \bm{a} } \Phi_{\bK}
~~~ \rightarrow~~~
\rme^{\rmi \hat{ \bm{\calK} }_R \cdot \bm{a} } \hat{\rho}_{\bK}^B \rme^{- \rmi \hat{ \bm{\calK} }_R \cdot \bm{a} }
 = \rme^{\rmi \bK \cdot \bm{a} } \hat{\rho}_{\bK}^B \rme^{- \rmi \bK \cdot \bm{a} } 
 = \hat{\rho}_{\bK}^B 
\,.
\eeq
This invariance of the baryon density matrix is inherited by the
one-body reduced density matrix. To see this, we write the total
magnetic translation operator as
\beq
T_{\bm a}
=
\prod_{j=1}^3 T_{\bm a}^{(j)},
\qquad
T_{\bm a}^{(j)}
=
\rme^{\rmi\hat{\bm{\calK}}_j\cdot\bm a},
\eeq
where $j=1,2,3$ denotes the particle slot.
For instance, the reduced density matrix for particle slot $1$ satisfies
\beq
\hat{\rho}_{\bK}^{(1)}
=
\Tr_{2,3}\big[\hat{\rho}_{\bK}^B\big]
=
\Tr_{2,3}\big[
T_{\bm a}\hat{\rho}_{\bK}^B T_{\bm a}^\dagger
\big]
=
T_{\bm a}^{(1)}
\Tr_{2,3}\big[\hat{\rho}_{\bK}^B\big]
T_{\bm a}^{(1)\dagger},
\eeq
where the unitary transformations acting on the traced-out particle
slots drop out by the invariance of the partial trace.
Thus,
\beq
T_{\bm a}^{(j)}
\hat{\rho}_{\bK}^{(j)}
T_{\bm a}^{(j)\dagger}
=
\hat{\rho}_{\bK}^{(j)} \,.
\eeq
Since the one-particle magnetic translation operator is diagonal in
flavor space, we may project this relation onto a definite flavor
$i=u,d$. Restricting further to the transverse sector, we obtain
\beq
T_{\bm a}^{i} \,
\hat{\rho}_{\bK_\perp}^{i}
T_{\bm a}^{i\dagger}
=
\hat{\rho}_{\bK_\perp}^{i},
\label{eq:comm_rho}
\eeq
where $T_{\bm a}^{i}$ is the one-particle magnetic translation
operator for a quark of charge $e_i$.

Since $\hat{\rho}_{\bK_\perp}^i$ commutes with every magnetic
translation $T_{\bm a}^i$, one might try to diagonalize the density
matrix together with the magnetic translations.
This is, however, not possible for all transverse translations
simultaneously, because the magnetic translation operators do not
commute with each other.
For a charged quark of flavor $i$, they satisfy
\beq
T_{\bm a}^i T_{\bm b}^i
=
\rme^{\rmi e_i \bm B\cdot(\bm a\times\bm b)}
T_{\bm b}^i T_{\bm a}^i .
\eeq
Therefore, one cannot construct a common eigenbasis of all transverse
magnetic translations. Instead, Eq.~\eqref{eq:comm_rho} constrains the
structure of the one-body density matrix through the irreducible
representation of the magnetic-translation group.

The transverse one-particle Hilbert space can be separated into two
independent sectors,
\beq
{\cal H}_\perp
=
{\cal H}_{\rm cyc}\otimes{\cal H}_{\rm gc} \,.
\eeq
Here ${\cal H}_{\rm gc}$ is the Hilbert space carrying an irreducible
representation of the magnetic-translation algebra, 
while ${\cal H}_{\rm cyc}$ describes the remaining cyclotron degree of
freedom, which is invariant under magnetic translations.
Hence magnetic translations may be written as
\beq
T_{\bm a}^{i}
=
{\bf 1}_{\rm cyc}\otimes T_{\bm a,{\rm gc}}^{i} \,.
\eeq
Since the set of operators $T_{\bm a,{\rm gc}}^i$ forms an irreducible
representation on the guiding-center Hilbert space, Schur's lemma,
together with Eq.~\eqref{eq:comm_rho}, implies
\beq
\hat\rho_{\bK_\perp}^i
=
\big[\hat\rho_{\bK_\perp}^i\big]_{\rm cyc}
\otimes {\bf 1}_{\rm gc} \,.
\eeq
Thus, each eigenvalue of the one-body density matrix is independent
of the guiding-center quantum number.
The nontrivial eigenvalue problem resides entirely in the cyclotron sector.

Diagonalizing the matrix $\hat\rho_{\bK_\perp}^i$
is still a nontrivial problem for a general $\bK_\perp$,
but at $\bK_\perp=0$ the problem is much simpler.
At $\bK_\perp=0$, we use the standard Landau basis
\beq
|n,\tilde n\ra_i
=
|n\ra_i^{\rm cyc}\otimes|\tilde n\ra_i^{\rm gc}.
\eeq
As shown below, this basis also diagonalizes $\hat\rho_{0_\perp}^i$.
Here, $n$ is the Landau-level (cyclotron) index, while $\tilde n$ labels the guiding-center states.
We then relate $\hat\rho_{\bK_\perp}^i$ and  $\hat\rho_{0_\perp}^i $ via a unitary transformation
\beq
\hat\rho_{\bK_\perp}^i
=
U_i(\bK_\perp)\,
\hat\rho_{0_\perp}^i\,
U_i^\dagger(\bK_\perp) \,,
\eeq
where $U_i(\bK_\perp)$ is the momentum-shift operator
\beq
U_i(\bK_\perp)
=
\exp\!\left[
\rmi \bm\kappa_\perp^i(\bK)\cdot\hat{\br}_\perp
\right] \,,
~~~~~~~~
\la \bm p_\perp | U_i (\bm K_\perp)
=
\exp\left[
- \bm\kappa_\perp^i (\bm K)\cdot
\frac{\partial}{\, \partial\bm p_\perp \,}
\right]
\la \bm p_\perp |
\,.
\eeq
Since $\hat{\br}_\perp$ contains both the cyclotron and guiding-center
coordinates, $U_i(\bK_\perp)$ acts on both sectors.
The explicit computations for the momentum shift are worked out in the next section.

Now we diagonalize $\rho_{0_\perp}^i$. First we note that the density matrix is diagonal for $\tilde{n}$,
\beq
\hat{ \rho}_{ 0_\perp}^i = \big[ \hat{\rho}_{ 0_\perp}^i \big]_{\rm cyc} \otimes {\bf 1}_{\rm gc}
 = \sum_{n,n',\tilde{n} }  \big[ \rho_{0_\perp}^i \big]^{nn'} |n, \tilde{n} \ra_i {}_i\la n', \tilde{n} |  \,.
\eeq
since ${\bf 1}_{\rm gc} = \sum_{\tilde{n}} | \tilde{n} \ra \la \tilde{n} |$.

The last step is to diagonalize the matrix 
$ \big[ \rho_{0_\perp}^i \big]^{nn'} = {}_i \la n, 0| \hat{\rho}_{ 0_\perp}^i |n',0 \ra_i $ where we chose $\tilde{n}=0$ as a representative.
In the symmetric gauge which we are taking, a state $|n,\tilde{n} \ra_i$ has the angular momentum $L_z = {\rm sgn} (e_i B) (\tilde{n} - n)$.
For fixed $\tilde n$, states with different $n$ carry different
angular momenta. Since $\hat\rho_{0_\perp}^i$ is rotationally
invariant, matrix elements between states with different $L_z$ must vanish.
Hence, the matrix is nonzero only for $n=n'$,
\beq
 \big[ \hat{ \rho}_{ 0_\perp}^i \big]^{nn'} 
 = \delta_{nn'} \big[ \hat{ \rho}_{ 0_\perp}^i \big]^{nn} 
 \equiv F_n^{i \perp} / S_\perp
 \,.
\eeq
Finally, defining $ |n,\tilde{n}; \bK_\perp \ra_i \equiv \rme^{\rmi \bm{\kappa}_\perp^i \cdot \hat{\br} } |n,\tilde{n} \ra_i$,
we obtain the spectral decomposition
\beq
\hat{ \rho }_{ \bK_\perp}^i 
 = \frac{1}{\, S_\perp \,}
 \sum_{n,\tilde{n} } F_{n}^{i \perp} \, 
 | n,\tilde{n}; \bK_\perp \ra_i {}_i \la n,\tilde{n}; \bK_\perp | \,.
\eeq
Here, we emphasize that $F_n^{i\perp}/S_\perp$ is an eigenvalue of the
one-body density matrix, whose corresponding eigenstate
$|n,\tilde n;\bK_\perp\ra_i$ depends on $\bK_\perp$ and
it is the superposition of various $| m, \tilde{m}\ra$ defined at $\bK_\perp=0$.
Hence, the quark occupation probability for the Landau bases, 
$\la n, \tilde{n} | \hat{ \rho }_{ \bK_\perp}^i | n, \tilde{n}\ra$,
does not correspond to $F_n^{i\perp}/S_\perp$;
for the former, we need to compute $\la n, \tilde{n} | m,\tilde m;\bK_\perp\ra_i = {}_i \la n,\tilde{n} | U_i (\bm K_\perp) |m,\tilde{m} \ra_i$,
and the matrix element manifestly depends on $\bK_\perp$.

The coefficient $F_{n}^{i \perp} $ can be explicitly computed as
\beq
 F_n^{i \perp}
  = S_\perp \int_{\br, \br'} \phi_{n,0}^{i*} (\br) {\rho}_{ 0_\perp}^i (\br, \br') \phi_{n,0}^i (\br') 
 = \frac{ 4\pi }{\, 4 A^i_\perp + |e_i B| \,} \bigg(\frac{\, 4A_\perp^i - |e_i B| \,}{\,  4A_\perp^i + |e_i B| \,} \bigg)^n 
 \,,
\eeq
where $\phi_{n,0}^{i} (\br) = \la \br | n, \tilde{n}=0 \ra_i$ is the $n$-th wave function.
In the large $B$ limit, only the $n=0$ component survives, 
as $A_\perp^i \sim |e_i B|/4 + O\big( (m\alpha)^2 /|B|^3 \big) $
where $O( 1/|B| )$ terms cancel.
The summation over $n$ reads
\beq
\sum_{\tilde{n} } 
\sum_{n}  \big[ \rho_{0_\perp}^i \big]^{nn} 
= \frac{\, |e_i B| \,} {\, 2\pi \,} S_\perp \sum_{n}  \big[ \rho_{0_\perp}^i \big]^{nn} 
= 1 
~~\rightarrow~~
\sum_{n} F_n^{i \perp} 
= \frac{\, 2\pi \,}{\, |e_i B| \,} \,.
\eeq
Here the guiding center index $\tilde{n}$ is limited by $S_\perp/( 2\pi/|e_i B| )$;
a single state occupies the area $2\pi/|e_i B| $ while the area of the system is $S_\perp$.

\subsubsection{ Quark distribution in a Landau basis }

Having diagonalized the density matrix, we can obtain the single-quark distribution in any chosen basis by projection.
We choose the momentum basis $|p_z\ra$ in the $z$ direction
and the Landau basis $|n,\tilde n\ra$ in the transverse plane.

For the longitudinal direction, the momentum distribution in a finite box is
\beq
{\cal F}^{\,i}_{p_z} (K_z)
\equiv
\la p_z | \hat{\rho}_{K_z}^{\,i} | p_z\ra
=
\frac{1}{\, L_z \,} F^i_{p_z;K_z} \,,
\eeq
which is proportional to the eigenvalue $F^i_{p_z;K_z}$.

For the transverse direction, 
the quark distribution in the Landau basis is
\beq
{\cal F}^{i}_{n \tilde{n} }(\bK_\perp)
\equiv
{}_i \la n,\tilde{n} | \hat\rho^i_{\bK_\perp} |n,\tilde{n} \ra_i
\,.
\eeq
Explicitly,
\beq
{\cal F}^{\,i}_{n\tilde n}(\bm K_\perp)
=
\frac{1}{\, S_\perp \,}
\sum_{m,\tilde{m} } F_m^{i \perp} \,
 \big| {}_i \la n, \tilde n | U_i (\bK_\perp) | m, \tilde{m} \ra_i \big|^2 \,.
 \label{eq:calF}
\eeq
The eigenvalues $F_m^{i \perp}$ themselves are independent of $\bm K_\perp$, 
and the $\bm K_\perp$-dependence appears through the unitary operator $U_i (\bK_\perp)$.
To evaluate the matrix element $ {}_i \la n, \tilde n | U_i (\bK_\perp) | m, \tilde{m} \ra_i$,
it is useful to decompose $\hat{\br}_i$ into the cyclotron ${\bm \eta}_i$ and guiding-center coordinates ${\bm X}_i$
\beq
\hat{\bm\eta}_i
=
\frac{\, \hat{\bm z} \times \hat{\bm\Pi}_i \,}{\, e_i B \,},
\qquad
\hat{\bm X}_i
=
\hat{\bm r}_{i\perp}-\hat{\bm\eta}_i .
\eeq
which act on ${\cal H}_i^{{\rm cyc}}$ and ${\cal H}_i^{{\rm gc}}$, respectively.
They satisfy
\beq
[\hat\eta_i^x,\hat\eta_i^y]
=
\frac{\rmi}{\, e_iB \,},
\qquad
[\hat X_i^x,\hat X_i^y]
=
-\frac{\rmi}{\, e_i B \,},
\qquad
[\hat X_i^a, \hat\eta_i^b]=0 .
\eeq
Thus, the cyclotron and guiding-center coordinates form two mutually commuting oscillator algebras.  
Since
\beq
U_i(\bm K_\perp)
=
\exp\left[
\rmi \bm{\kappa}_\perp^i \cdot \hat{\br}_{i\perp}
\right],
\eeq
the decomposition
$\hat{\br}_{i\perp} = \hat{\bm X}_i + \hat{\bm\eta}_i$
gives
\beq
U_i
=
\exp\left[
\rmi {\bm \kappa}_\perp^i \cdot\hat{\bm\eta}_i
\right]
\exp\left[
\rmi {\bm \kappa}_\perp^i \cdot\hat{\bm X}_i
\right] \,,
\eeq
we find
\beq
{}_i\la n,\tilde n|
U_i
|m,\tilde m\ra_i
=
\la n|
\rme^{\rmi \bm{\kappa}_\perp^i \cdot \hat{\bm\eta}_i}
|m \ra_i^{{\rm cyc}}
\,
\la \tilde n|
\rme^{\rmi \bm{\kappa}_\perp^i \cdot \hat{\bm X}_i}
| \tilde{m} \ra_i^{ {\rm gc} } \,.
\eeq
Actually, for computations of the probability Eq.~\eqref{eq:calF},
it is not necessary to evaluate
$\la \tilde n| \rme^{\rmi \bm{\kappa}_\perp^i \cdot \hat{\bm X}_i} | \tilde{m} \ra_i^{ {\rm gc} } $.
Noting that the weight $ F_m^{i\perp} $ is independent of the guiding-center quantum number $\tilde m$,
we use the completeness in the guiding-center sector,
$\sum_{\tilde m}|\tilde m\ra_i^{\rm gc} \, {}_i^{\rm gc}\la\tilde m|={\bf 1}_{\rm gc}$,
together with the unitarity of
$\exp[\rmi\bm\kappa_\perp^i\cdot\hat{\bm X}_i]$.
We find
\beq
{\cal F}^{\,i}_{n\tilde n}(\bm K_\perp)
=
\frac{1}{\, S_\perp \,}
\sum_{m }
F_m^{i \perp}
 \big| {}_i \la n | \rme^{\rmi \bm{\kappa}_\perp^i \cdot \hat{\bm\eta}_i} | m \ra_i \big|^2 \,.
 \label{eq:calF2}
\eeq
To compute the matrix element, we note that
the ${\bm \eta} \sim \hat{z} \times {\bm \Pi}$ can be expressed in terms of the creation and annihilation operators
which raise and lower the quantum number $n$ as $\hat{a}_i |n \ra_i = \sqrt{n} |n-1\ra_i $ and $\hat{a}_i^\dag |n\ra_i = \sqrt{n+1} |n+1\ra_i$.
The exponent of $\rme^{\rmi \bm{\kappa}_\perp^i \cdot \hat{\bm \eta}_i }$ can be written as
[$s_i \equiv {\rm sgn} (e_iB)$]
\beq
\rmi \bm{\kappa}_\perp^i \cdot \hat{\bm\eta}_i
= \alpha_i \hat{a}_i^\dag - \alpha_i^* \hat{a}_i \,,
~~~~~~~~
\alpha_i 
= \frac{\, \rmi \kappa_x^i - s_i \kappa_y^i  \,}{\, \sqrt{ 2 | e_i B | }\,} \,.
\eeq
We define
\beq
\calW_{nm} (\bK_\perp) 
\equiv 
\big| {}_i \la n | \rme^{\rmi \bm{\kappa}_\perp^i \cdot \hat{\bm\eta}_i} | m \ra_i \big|^2
= \rme^{-x_i} \frac{\, {\rm min} [n,m] !\,}{\, {\rm max} [n,m] !\,} \,
 x_i^{ |n-m| } \big[ L^{ |n-m| }_{ {\rm min} (n,m) } (x_i) \big]^2
 \,,
\eeq
where $L_n^\alpha $ is the associated Laguerre polynomial and $ x_i \equiv (\kappa^i_\perp)^2 / 2|e_i B|$ as a function of $K_\perp$.

\begin{figure}[tb]
\centering
\includegraphics[width=0.95\linewidth]{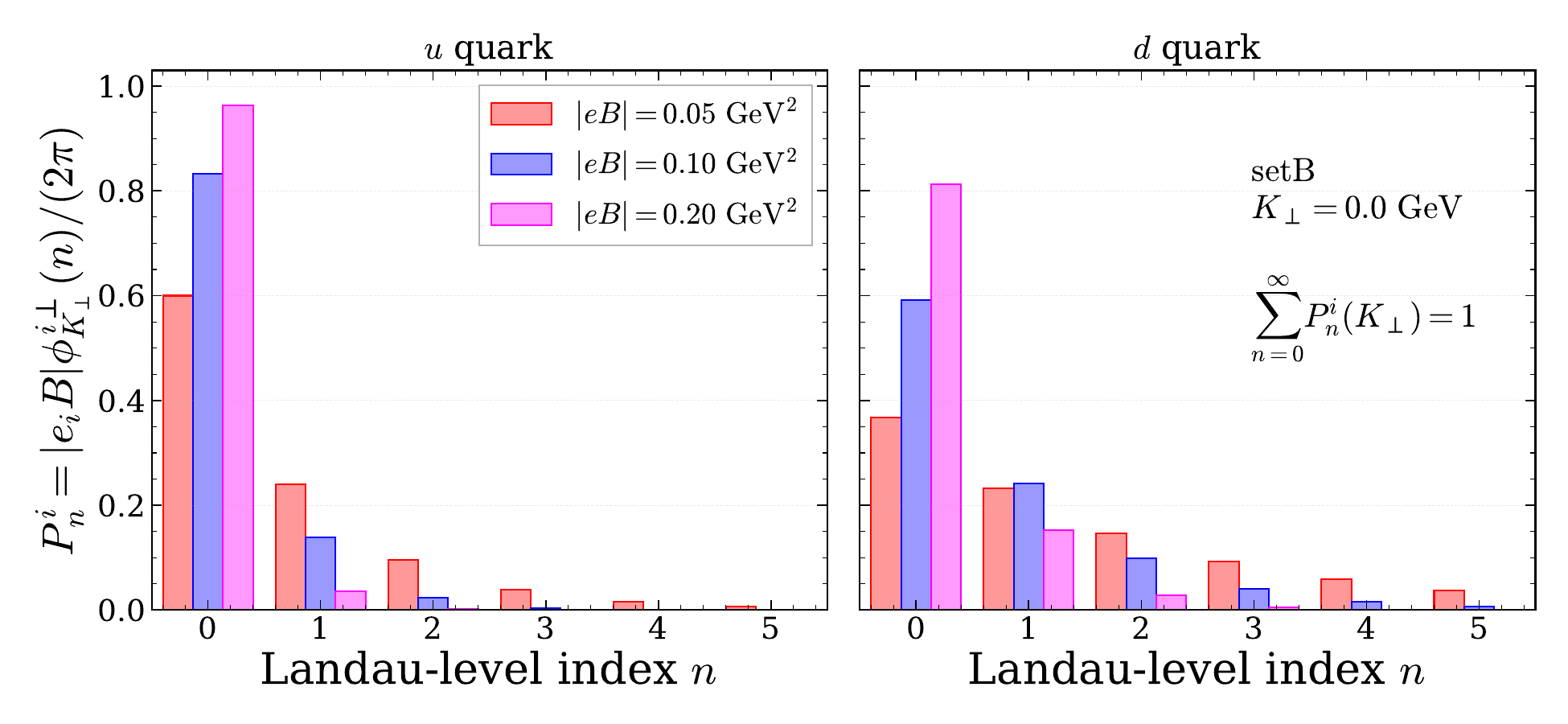}
\caption{Landau-level distribution of a single quark in the lowest neutral-baryon state for parameter set B at $K_\perp=0$. 
The left and right panels show the $u$- and $d$-quark distributions, respectively. 
We plot $P_n^i=|e_iB|\phi_{K_\perp}^{i\perp}(n)/2\pi$ for $|eB|=0.05$, $0.10$, and $0.20\,{\rm GeV}^2$. 
As the magnetic field increases, the distribution becomes increasingly concentrated in the lowest Landau level.
}
\label{fig:phi_perp_Landau_hist}
\end{figure}
\begin{figure}[tb]
\centering
\includegraphics[width=1.0\linewidth]{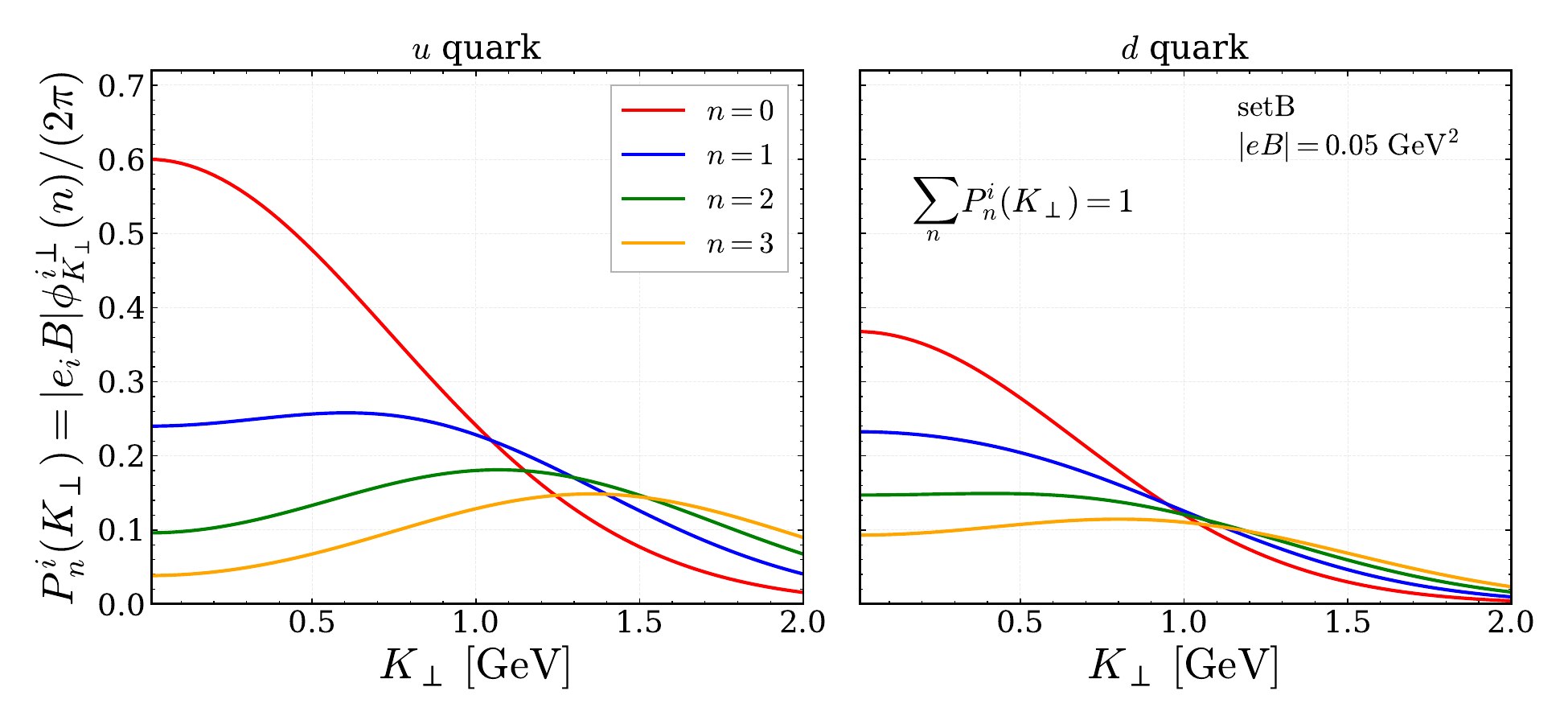}
\caption{Dependence of the single-quark Landau-level occupation probabilities
on the transverse baryon pseudomomentum $K_\perp$ for parameter set B at $|eB|=0.05\,{\rm GeV}^2$. 
The left and right panels show the $u$- and $d$-quark distributions, respectively. 
We plot
$P_n^i(K_\perp)=|e_iB|\phi_{K_\perp}^{i\perp}(n)/2\pi$
for the Landau levels $n=0,1,2,$ and $3$.
As $K_\perp$ increases, the occupation probability is redistributed from the
lowest Landau level to higher Landau levels.
}
\label{fig:phi_perp_vs_Kperp_set2_B0050}
\end{figure}

The transverse distribution in the fixed Landau basis then takes the form
\beq
{\cal F}^{\,i}_{n \tilde{n} }(\bm K_\perp)
=
\frac{1}{\, S_\perp \,} \sum_{m } F_m^{i \perp} \calW_{nm} (\bK_\perp) 
\equiv
\frac{\, \varphi_{\bK_\perp}^{i \perp} (n) \,}{\, S_\perp \,}
\,.
\eeq
where ${\cal F}^{\,i}_{n\tilde n}(\bm K_\perp)$ is independent of $\tilde{n}$.
Figure~\ref{fig:phi_perp_Landau_hist} illustrates the transverse single-quark distribution
for parameter set B at $K_\perp=0$. As the magnetic field
increases, the distribution becomes increasingly concentrated in the lowest Landau level. 
The approach to the lowest-Landau-level regime differs between the
$u$ and $d$ quarks because their transverse density matrices involve
different charges and widths.

The dependence on the baryon transverse pseudomomentum is illustrated in
Fig.~\ref{fig:phi_perp_vs_Kperp_set2_B0050} for $|eB|=0.05\,{\rm GeV}^2$. 
As $K_\perp$ increases, the unitary shift $U_i(K_\perp)$ mixes different
Landau levels, and the probability is gradually redistributed from the lowest Landau level to higher levels. 
The detailed redistribution differs between the $u$ and $d$ quarks
because the overlap factor $W_{nm}(K_\perp)$ depends on the quark charge
through both $\kappa_\perp^i$ and $|e_iB|$.

Since the transverse and longitudinal density matrices factorize, the single-quark distribution in a baryon with pseudomomentum $\bm K$ is
\beq
\varphi^{i}_{\bK} (n, p_z)
=
\varphi^{i \perp}_{ \bK_\perp } (n) \, \varphi^{i} (p_z-K_z/3)
\,.
\eeq
The volume-independent distribution is normalized with the
Landau-level phase-space measure as
\beq
\frac{\, |e_i B| \,}{2\pi}
\sum_n \int \frac{\, \rmd p_z \,}{ 2\pi }\,
\varphi^{i}_{\bK} (n, p_z) =1 \,,
\label{eq:normalization_phi}
\eeq
where we used
\beq
\sum_n W_{nm} (K_\perp)=1 \,,~~~~~~~~
\sum_m F_m^{i \perp} = \frac{\, 2\pi \,}{\, |e_i B| \,} \,,~~~~~~~~
\int_{p_z} \, \varphi^i (p_z - K_z/3) =1 \,.
\eeq
\begin{figure}[tb]
\centering
\includegraphics[width=0.55\linewidth]{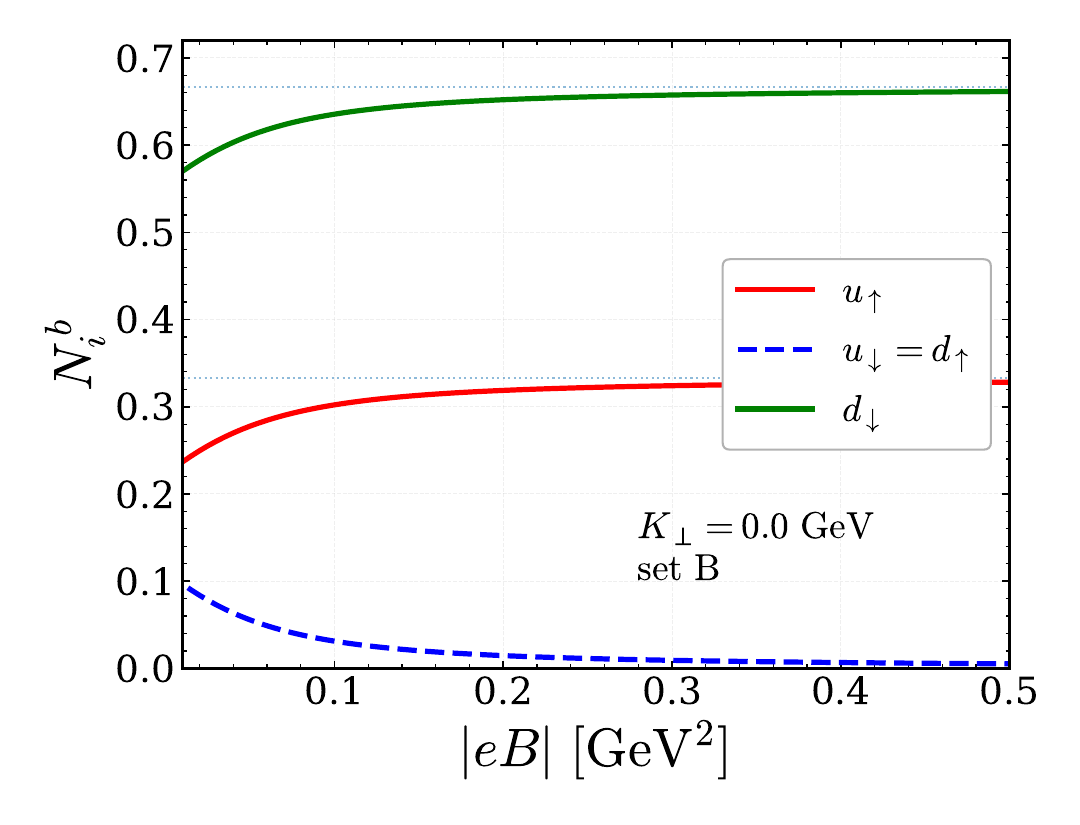}
\caption{
Spin-flavor baryon-number fractions $N_i^b$ in the lowest
$\widetilde n_\downarrow$ branch as functions of the magnetic field
for parameter set B at $\bK=0$.
At $B=0$, the fractions are
$N_{u_\uparrow}=2/9$,
$N_{u_\downarrow}=N_{d_\uparrow}=1/9$, and
$N_{d_\downarrow}=5/9$.
With increasing $|eB|$, the state approaches the strongly polarized
configuration $d_\downarrow d_\downarrow u_\uparrow$, for which
$N_{u_\uparrow}\to1/3$,
$N_{u_\downarrow}=N_{d_\uparrow}\to0$, and
$N_{d_\downarrow}\to2/3$.
}
\label{fig:spin_flavor_weights_setB_Kperp_0}
\end{figure}

\section{Baryon and quark occupation probabilities} \label{sec:dense}

\subsection{ Sum rules }

We now construct the quark occupation probabilities in baryonic matter.
We denote the baryon eigenstates by $b$ in which the spin-flavor and orbital quantum numbers are all distinguished,
the quark spin-flavor species by $i =(u_\up, u_\down, d_\up, d_\down, \cdots)$, and the quark Landau-level index by $n$.
We emphasize that, at finite $B$, the usual baryon classification, $(n_\up, p_\up,\cdots)$ and so on, no longer holds. 
The single-quark distribution inside a baryon, derived in Sec.~3, is
denoted by
\beq
\varphi^{i/b}_{\bK} (n,p_z) \,,
\eeq
where $\bm K=(\bm K_\perp,K_z)$ is the baryon (pseudo)momentum.
The occupation probability of a baryon state $(b,\bm K)$ is denoted by
$f_{b }(\bm K)$.

The spin-flavor factors $N_i^b(\bK)$ entering the convolution are
determined by the spin composition of the baryon state discussed in
Sec.~\ref{sec:spin-flavor}.  Figure~\ref{fig:spin_flavor_weights_setB_Kperp_0} illustrates these
factors for the lowest $\widetilde n_\downarrow$ branch at $\bK=0$
for parameter set B.
At $B=0$, the neutron spin-flavor wave function gives
\beq
N_{u_\uparrow}=\frac{2}{9},\qquad
N_{u_\downarrow}=N_{d_\uparrow}=\frac{1}{9},\qquad
N_{d_\downarrow}=\frac{5}{9}.
\eeq
As the magnetic field increases, the baryon becomes increasingly
polarized toward the
$d_\downarrow d_\downarrow u_\uparrow$ configuration.
Consequently,
\beq
N_{u_\uparrow}\to\frac{1}{3},\qquad
N_{u_\downarrow},\,N_{d_\uparrow}\to0,\qquad
N_{d_\downarrow}\to\frac{2}{3},
\qquad B\to\infty .
\eeq
This spin-flavor redistribution is important for quark saturation,
because it enhances the weight carried by the $d_\downarrow$ states in the lowest baryon branch.

The quark occupation probability is then obtained by convoluting the
baryon distribution with the single-quark distribution 
\beq
f_n^{i} (p_z)
=
\sum_{b}
 \int_{\bK} N_{i}^{b } (\bK)
f_{b }(\bm K)\,
\varphi^{ i/b }_{\bK} (n,p_z)
 \,.
\label{eq:quark_occ_general}
\eeq
Since the baryons are color singlets, the three colors contribute equally;
for a given color, $ f_n^{i}  \equiv [f_n^{i} ]^R =  [f_n^{i} ]^G =  [f_n^{i} ]^B $.
Here, $N_{i}^{b }$ denotes the baryon number ($1/3$ of quark number) of quarks of species $i$
contained in the baryon state $b$.
For instance, in a neutron state at $B=0$, 
the baryon-number fraction carried by each quark species is 
$N^{u_\up}_{n_\up} + N^{u_\down}_{n_\up} = 1/3$ and $N^{d_\up}_{n_\up} + N^{d_\down}_{n_\up} = 2/3$, 
since a neutron with a given spin contains one $u$ quark and two $d$ quarks.
Indeed, integrating the quark distribution for a given color leads to the total baryon number,
$n_{\rmB} = \sum_i n_{i}^R = \sum_i n_{i}^G = \sum_i n_{i}^B$, 
\beq
n_{\rmB} 
= \sum_{i} \frac{\, |e_{i} B| \,}{2\pi}
\sum_n \int_{p_z} f_n^{i} (p_z)
= \sum_{i} \sum_{b}
 \int_{\bK} N_{i}^{b } (\bK) f_{b}(\bm K)
= \sum_{b} n_{ b } 
\,
\eeq
where we have used the normalization condition Eq.~\eqref{eq:normalization_phi} for $\varphi^{i/b}_{\bK}$
and $\sum_{i} N_{i}^{b } (\bK) = 1$ for any baryon state $b$ and pseudomomentum $\bK$.

At zero temperature, the baryon states are filled from the lowest energies
up to a maximum energy $E_{\rm max}$,
\beq
f_{b }(\bm K) = \Theta \big( E_{\rm max} - E_{b }(\bm K) \big) \,.
\label{eq:baryon_occ}
\eeq
For an ideal baryon gas before quark saturation occurs,
$E_{\rm max}=\mu_B$.


\subsection{ Ideal baryon gas }

\begin{figure}[tb]
\centering
\includegraphics[width=0.50\linewidth]{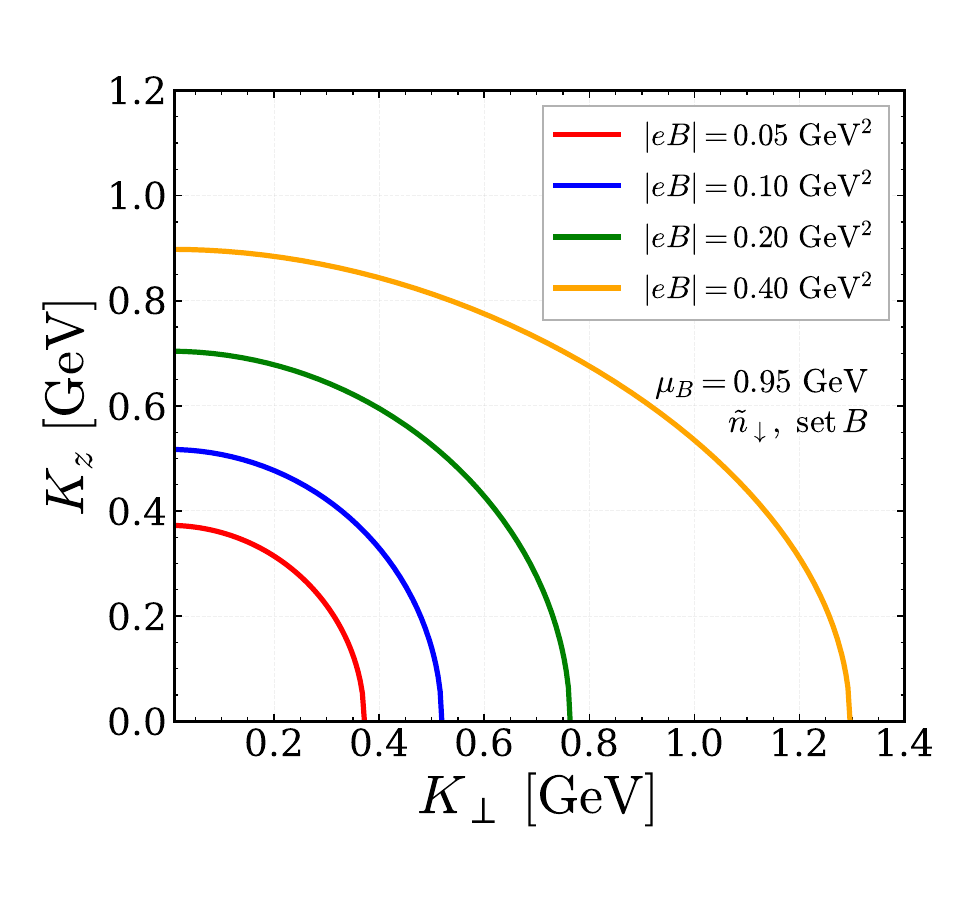}
\caption{
Fermi-surface cross sections of the ground state baryon,
lowest \(\widetilde n_\downarrow\) branch, in the $(K_\perp,K_z)$ plane at $\mu_B=0.95$ GeV for several magnetic fields.
With increasing $|eB|$, the occupied region expands as the baryon energy is lowered.
The transverse flattening of the single-baryon dispersion produces an increasingly anisotropic Fermi surface elongated in the $K_\perp$ direction.
Results are for parameter set B.
}
\label{fig:fermi_surface_setB_muB095}
\end{figure}
\begin{figure}[tb]
\centering
\includegraphics[width=0.495\linewidth]{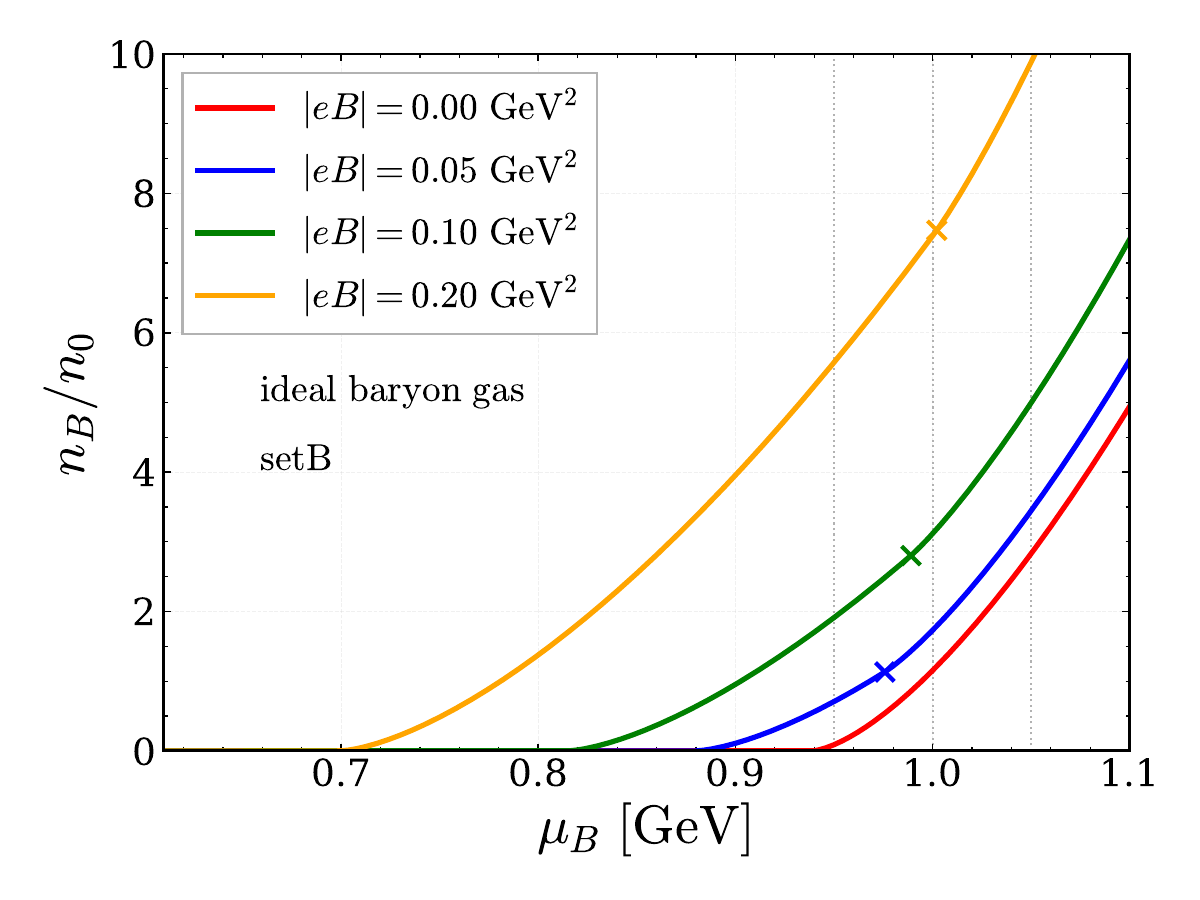}
\includegraphics[width=0.495\linewidth]{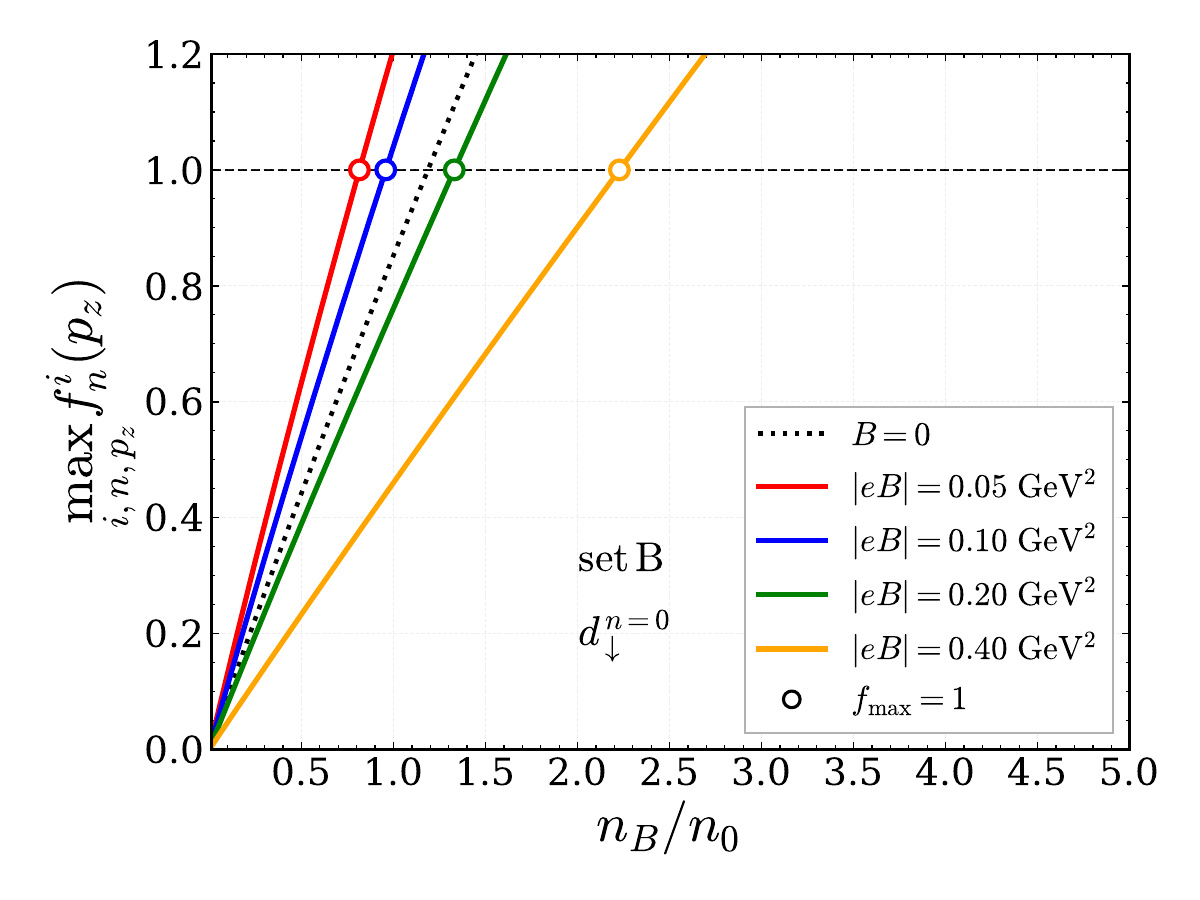}
\caption{
(Left panel) Baryon density $n_B$ as a function of the baryon chemical
potential $\mu_B$ for an ideal gas of charge-neutral baryons at several magnetic fields.
The magnetic field lowers the onset chemical potential of the lowest
$\widetilde n_\downarrow$ branch and enhances the baryon density at
fixed $\mu_B$.
Crosses mark the onset of the $\widetilde n_\uparrow$ branch.
(Right panel) Maximum quark occupation probability,
$\max_{i,n,p_z}f_n^i(p_z)$, as a function of the baryon density.
At finite $B$, the maximum occurs in the $d_\downarrow$, $n=0$ state.
Open circles indicate the quark-saturation density.
Results are shown for parameter set B, with
$n_0=0.16~{\rm fm}^{-3}$.
}
\label{fig:ideal_gas_setB}
\end{figure}

Below we consider an ideal baryon gas for which
$E_{\rm max}=\mu_B$.
For each baryon branch $b$, the occupied region in pseudomomentum
space is determined by
\beq
E_b(K_\perp,K_z)\leq\mu_B .
\eeq
We use this occupied phase space to evaluate the baryon density
and the corresponding quark occupation probabilities.
We define $K_{\perp \rm{max} }^{b }$ by
\beq
E_{b}
\big( K_{\perp \, \rm{max} }^{b} \,, 0 \big)
=
\mu_B \,,
\eeq
and, for a fixed $K_\perp$, the maximum of the longitudinal momentum $ K_{z,\rm max}^{b} (K_\perp)$ by
\beq
E_{b}
\big( K_\perp, K_{z, \rm max}^{b}(K_\perp) \big) = \mu_B \,.
\eeq
Accordingly, Eq.~\eqref{eq:quark_occ_general} can be written as
\beq
f_n^{i} (p_z)
=
\sum_{b}
\int_0^{K_{\perp,\rm max}^{b }}
\frac{\, K_\perp \rmd K_\perp \,}{2\pi}
\int_{ -K_{z,\rm max}^{b }(K_\perp) }^{ K_{z,\rm max}^{b }(K_\perp) }
\frac{\, \rmd K_z \,}{2\pi}\,
N_{i}^{b } (\bK)
\varphi^{i/b}_{\bK} (n,p_z)
 \,.
\label{eq:quark_occ_explicit}
\eeq
The baryon number density is
\beq
n_{\rmB} (\mu_B)
=
2\sum_{b}
\int_0^{K_{\perp,\rm max}^{b}}
\frac{\, K_\perp \rmd K_\perp \,}{2\pi}
\int_0^{K_{z,\rm max}^{b}(K_\perp)}
\frac{\, \rmd K_z \,}{2\pi} \,.
\label{eq:baryon_density}
\eeq
%

\subsection{ Quark saturation in magnetic fields }
\label{sec:quark_saturation}

We now examine when the quark occupation probability constructed above reaches unity, 
its maximum possible value.
For a given baryon density and magnetic field, 
we define the maximum single-quark occupation probability by
\beq
f_{\rm max}(n_B,B)
\equiv
\max_{i,n,p_z} f_n^i(p_z) \,.
\label{eq:fmax}
\eeq
The quark-saturation density $n_B^{\rm qsat}$ is then defined by
\beq
f_{\rm max}\bigl(n_B^{\rm qsat},B\bigr)=1 \,.
\label{eq:nB_qsat}
\eeq
Below this density, all quark one-particle states have occupation
probabilities smaller than unity.
Beyond this point, continuing the same baryonic filling would make
the occupation of some quark states exceed unity.
We therefore use $n_B^{\rm qsat}$ as an estimate of the upper limit
of applicability of the purely baryonic description, without making
any assumption about the state of matter beyond this point.

\begin{figure}[tb]
\centering
\includegraphics[width=0.6\linewidth]{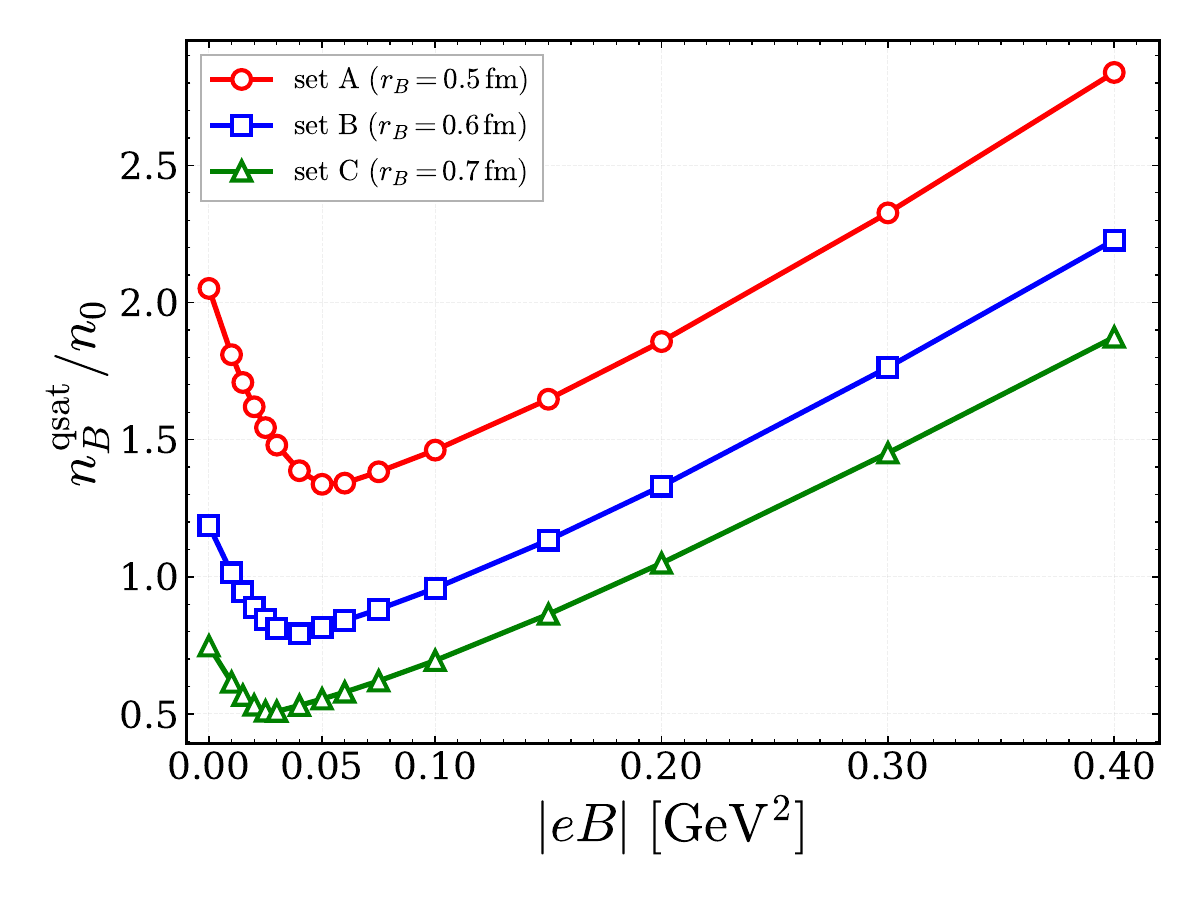}
\caption{
Quark-saturation density $n_B^{\rm qsat}$ as a function of the magnetic
field for parameter sets A--C.
The saturation density is defined by
$f_{\rm max}(n_B^{\rm qsat},B)=1$, where
$f_{\rm max}=\max_{i,n,p_z}f_n^i(p_z)$.
For all three parameter sets, a weak magnetic field initially lowers
the saturation density, while stronger fields increase it again,
leading to a characteristic nonmonotonic dependence on $|eB|$.
The variation among the three curves reflects the sensitivity to the
zero-field baryon size.
}
\label{fig:qsat_B}
\end{figure}

The right panel of Fig.~\ref{fig:ideal_gas_setB} shows $f_{\rm max}$
as a function of $n_B$ for parameter set B.
At finite $B$, the $d_\downarrow$, $n=0$ state reaches saturation first.
Figure~\ref{fig:qsat_B} summarizes the resulting saturation density for
parameter sets A--C.
For all three sets, $n_B^{\rm qsat}$ first decreases at weak magnetic
fields and then increases again at stronger fields.

The turnover reflects a competition between two magnetic-field effects.
At relatively weak fields, the spin-flavor polarization of the baryon
and the concentration of quarks in the lowest Landau levels
enhance the $d_\downarrow$ occupation, leading to quark saturation at
lower density.
At stronger magnetic fields, however, the internal quark wave function
is compressed in the transverse directions, as shown in
Fig.~\ref{fig:baryon_mass_B}, 
suggesting  reduced overlap between quark wave functions in neighboring baryons 
and hence a tendency to delay quark saturation.
More precisely, the number of available guiding-center states per unit
transverse area increases with $|eB|$, since each guiding-center state
occupies a spatial area of order $1/|eB|$.
Consequently, at a given baryon density, the occupation probability of
the lowest cyclotron state at each guiding center increases more slowly,
delaying quark saturation.
These competing effects, spin-flavor polarization and spatial squeezing,
produce a minimum in $n_B^{\rm qsat}$.

\section{ Summary } \label{sec:summary}

We have studied the onset of quark saturation in magnetized matter
composed of electrically neutral $ddu$ baryons.
Using a simple constituent-quark model, we first examined the spectra
and internal structure of the low-lying neutral baryons in a magnetic field.
The magnetic field strongly modifies both the spin-flavor composition
and the orbital wave function.
In particular, the lowest $\widetilde n_\downarrow$ branch becomes
increasingly polarized toward the
$d_\downarrow d_\downarrow u_\uparrow$ configuration.
At the same time, the baryon is squeezed in the transverse direction,
and its transverse dispersion becomes strongly flattened.
The resulting large transverse effective mass indicates an effective
dimensional reduction of the baryon dynamics.

Starting from these baryon wave functions, we constructed the one-body
reduced density matrix and derived the single-quark distributions.
The longitudinal distribution is naturally described in momentum space,
while the transverse distribution is expressed in the Landau-level basis.
The magnetic field concentrates the quark distribution toward the
lowest Landau levels, while the increasing guiding-center degeneracy
reduces the occupation probability of each transverse one-particle state.
Finite transverse baryon pseudomomentum further redistributes the
occupation among different Landau levels.

We then considered an ideal gas of the neutral baryons and evaluated
the quark occupation probabilities as the baryon density is increased.
Quark saturation is defined by the point at which the maximum
single-quark occupation probability reaches unity.
At finite magnetic field, the $d_\downarrow$ quark in the lowest
Landau level is the first state to reach saturation.
The resulting saturation density shows a characteristic nonmonotonic
dependence on the magnetic field: it first decreases at weak fields
and then increases again at stronger fields.
This behavior is found for all parameter sets considered here.

The nonmonotonic behavior reflects a competition between spin-flavor
polarization and spatial squeezing.
At relatively weak fields, the enhanced $d_\downarrow$ component
promotes quark saturation.
At stronger fields, 
transverse compression increases the number of available guiding-center states per unit area 
and thereby reduces the occupation probability of each transverse one-particle state
and delays saturation.
Thus, magnetic fields can either reduce or extend the density range
over which a purely baryonic description remains applicable.
The present analysis does not specify the state of matter beyond quark
saturation; determining how quark-level constraints modify dense matter
beyond this point is left for future work.

Several extensions of the present framework are worth pursuing. 
For applications to realistic neutron-star matter, 
it will be necessary to include additional baryonic species together with leptons under charge-neutrality and beta-equilibrium conditions, 
and to go beyond the ideal baryon gas by incorporating realistic baryon-baryon interactions.
The inclusion of charged baryons requires an additional extension, 
since their transverse center-of-mass motion is itself Landau quantized 
and cannot be characterized by the commuting pseudomomentum components used for the neutral baryons considered here.

Finite-temperature effects require a more substantial extension. 
In addition to thermal smearing of the occupation probabilities, thermally excited baryons, mesonic excitations, particle-hole excitations, 
and higher Landau levels may become important. 
A consistent treatment of such excited states will likely require a relativistic formulation \cite{Andreichikov:2014PRD,Andreichikov:2019},
since the magnetic-field dependence of states involving higher Landau levels of quarks
can differ qualitatively from that of the low-lying states considered in the present nonrelativistic framework.
Developing such a relativistic finite-temperature description is therefore an important direction for future work.

Another interesting direction is to test the present mechanism in theories accessible to first-principles calculations at finite density. Two-color QCD provides one such setting, where the absence of the sign problem can be maintained even in an external magnetic field for suitable choices of the quark electric charges \cite{Kogut:2000ek,Brauner:2019rjg}.
Recent lattice studies have explored its finite-density phase structure, equation of state, and hadron spectrum \cite{Iida:2019rah,Iida:2022hyy,Iida:2024irv,Iida:2026qc2d}.
QCD at finite isospin chemical potential provides another useful laboratory without a sign problem at vanishing magnetic field \cite{Son:2000xc,Brandt:2017oyy,Brandt:2022hwy,Abbott:2023coj,Abbott:2024zms},
although the introduction of an external magnetic field with physical quark charges generally requires additional care \cite{Endrodi:2014lja}. 

Complementarily, substantial progress has recently been made in dense QCD in $(1+1)$ dimensions using tensor-network methods. In these calculations, quark momentum distributions and the formation of quark Fermi seas can be studied directly \cite{Hayata:2024jxt,Fujikura:2026qcd2}. Such lower-dimensional theories may therefore provide a particularly useful setting for investigating the microscopic relation between quark occupation, baryonic degrees of freedom, and quark saturation.

\appendix
\setcounter{section}{0}

\section{Matrix elements}\label{app:matrix_elements}

For the lowest orbital state, the expectation values of the
short-range interactions can be reduced to one-dimensional
integrals.  For the $dd$ pair,
\beq
\bm r_{12}=-\bm\rho ,
\eeq
while for the two $ud$ pairs,
\beq
\bm r_{13}
=
-\bm\lambda-\frac{\bm\rho}{2},
\qquad
\bm r_{23}
=
-\bm\lambda+\frac{\bm\rho}{2}.
\eeq
The two $ud$ matrix elements are identical because of the
$\bm\rho\to-\bm\rho$ symmetry.
The transverse displacement relevant for the $ud$ pairs is
\beq
\lambda_0^2
\equiv
| \bm\lambda_\perp^0 |^2
=
\frac{\, B_q^2K_\perp^2 \,}{\, 4(B_q^2+3m\alpha)^2 \,} \,.
\label{eq:lambda0_app}
\eeq
%

\subsection{Color-electric interaction}

For the color-electric interaction, we use
\beq
\frac{1}{\, r \,}
=
\frac{1}{\, \sqrt{\pi} \,}
\int_0^\infty
\frac{ds}{\, \sqrt{s} \,}\,
e^{-sr^2},
\eeq
and define
\beq
{\cal Y}_E (a,c,r_0)
\equiv
\frac{1}{\, \sqrt{\pi} \,}
\int_0^\infty
\frac{ds}{\, \sqrt{s} \,}\,
\frac{1}{\, (1+as)\sqrt{1+cs} \,}
\exp\left[
-\dfrac{s r_0^2}{1+as}
\right] \,.
\label{eq:YE_function_app}
\eeq
The color-electric matrix elements are
\begin{align}
\left\langle v^E_{dd}\right\rangle_0^K
&=
-\frac{2}{\, 3\,} \alpha_s(Q_{dd})
\calY_E
\left(
\beta_{dd}^\perp\,,
\beta_{dd}^z\,,
0
\right),
\notag \\
\left\langle v^E_{ud}\right\rangle_0^K
&=
-\frac{2}{\, 3 \,}\alpha_s(Q_{ud})
\calY_E
\left(
\beta_{ud}^\perp\,,
\beta_{ud}^z\,,
\lambda_0
\right) \,.
\label{eq:VEud_app}
\end{align}
For vanishing transverse displacement,
\beq
\calY_E (a,c,0)
=
\frac{2}{\sqrt{\pi a}}
\frac{\, \sin^{-1}\sqrt{1-c/a} \,}{ \sqrt{1-c/a} } \,,
\qquad a > c \,,
\eeq
while
\beq
{\cal Y}_E (a,c,0)
=
\frac{2}{\sqrt{\pi a}}
\frac{\, \sinh^{-1}\sqrt{c/a-1} \,}{ \sqrt{c/a-1} } \,,
\qquad c>a.
\eeq
At $a=c$,
\beq
{\cal Y}_E (a,a,0)
=
\frac{2}{\, \sqrt{\pi a} \,} \,.
\eeq
%
\subsection{Color-magnetic interaction}

For the color-magnetic interaction, we relativize the standard
factor $\alpha_s/(m_i m_j)$ by replacing
\beq
\frac{\alpha_s}{m_i m_j}
\ \longrightarrow\
\frac{\alpha_s}{E_iE_j}.
\eeq
Taking $|\bm p_1|=|\bm p_2|=|\bm q|/2$ and
$m_i=m_j=m$, we use
\beq
\widetilde v^M_{ij}(\bm q)
=
C_M
\frac{\, 4\pi \,}{9}\,
\frac{\, \alpha_s(Q_{ij}) \,}{\, m^2+\bm q^2/4 \,} \,,
\label{eq:VM_momentum_app}
\eeq
where the Fourier-transform convention is
\beq
v(\bm r)
=
\int\frac{\, \rmd^3 \bq \,}{(2\pi)^3}\,
\rme^{\rmi \bm q\cdot\bm r}\,
\widetilde v(\bm q).
\eeq
The corresponding coordinate-space interaction is
\beq
v^M_{ij}(r)
=
C_M \frac{\,  4 \alpha_s (Q_{ij}) \,}{\, 9 \,}
\frac{\, \rme^{-2mr} \,}{ r }.
\label{eq:VM_coordinate_app}
\eeq
Using a formula
\beq
\frac{\, \rme^{-\mu r} \,}{r}
=
\frac{1}{\, \sqrt{\pi} \,}
\int_0^\infty
\frac{\, \rmd s \,}{\, \sqrt{s} \,}\,
\exp\left(
- sr^2 - \frac{\, \mu^2 \,}{\, 4s \,}
\right) \,,
\label{eq:Yukawa_Schwinger}
\eeq
we define
\beq
\begin{split}
{\cal Y}_M (a,c,r_0;\mu)
\equiv{}&
\frac{1}{\sqrt{\pi}}
\int_0^\infty
\frac{ds}{\sqrt{s}}\,
\frac{1}{(1+as)\sqrt{1+cs}}
\times
\exp\left[
-\frac{\, \mu^2 \,}{\, 4s \,}
-\frac{\, s r_0^2 \,}{\, 1 + as \,}
\right].
\end{split}
\label{eq:Y_function}
\eeq
The magnetic matrix elements are then
\begin{align}
\left\langle v^M_{dd}\right\rangle_0^K
&=
C_M \frac{\,  4 \alpha_s (Q_{ij}) \,}{\, 9 \,}
{\cal Y}_M
\left(
\beta_{dd}^\perp\,,
\beta_{dd}^z\,,
0\,;2m
\right),
\notag \\
\left\langle v^M_{ud}\right\rangle_0^K
&=
C_M \frac{\,  4 \alpha_s (Q_{ij}) \,}{\, 9 \,}
{\cal Y}_M
\left(
\beta_{ud}^\perp \,,
\beta_{ud}^z \,,
\lambda_0\,; 2m
\right).
\label{eq:VMddud_app}
\end{align}
For $a=c$ and $r_0=0$, the integral can be evaluated explicitly,
\beq
{\cal Y}_M (a,a,0;\mu)
=
\frac{2}{\sqrt{\pi a}}
-
\mu
\exp\left(\frac{\mu^2a}{4}\right)
\operatorname{erfc}
\left(
\frac{\, \mu\sqrt{a} \,}{2}
\right).
\label{eq:Y_isotropic}
\eeq

\clearpage

\isPreprints{}

%

\funding{
This work is supported by JSPS KAKENHI Grant No. 26K07077.
}

\dataavailability{ Data are available upon request. } 

\conflictsofinterest{The authors declare no conflicts of interest.} 


%

\acknowledgments{
We thank Drs. Akinobu Dote, GuangJuan Wang, Asanosuke Jinno, and Mr. Haru Yoshimori for useful discussions at the KEK hadron meeting.
}

\reftitle{References}
\bibliography{ref}

\isPreprints{}

\end{document}